\documentclass[twocolumn]{aastex631}
\usepackage{amsmath}
\usepackage[utf8]{inputenc}
\usepackage{placeins}

\begin{document}



\title{Discovery of a New Spectral Transition in Swift\,J0243.6+6124 in the Sub-Eddington Regime}
\author[0009-0008-0518-6795]{\textbf{Bo-Yan Chen}}
\affiliation{\textbf{State Key Laboratory of Particle Astrophysics, Institute of High Energy Physics, Chinese Academy of Sciences, Beijing 100049, People's Republic of China}}
\affiliation{\textbf{University of Chinese Academy of Sciences, Chinese Academy of Sciences, 
100049 Beijing, People's Republic of China}}
\email{bychen@ihep.ac.cn}

\correspondingauthor{\textbf{Shu Zhang}}
\email{szhang@ihep.ac.cn}
\author[0000-0003-3972-2564]{\textbf{Shu Zhang}}
\affiliation{\textbf{State Key Laboratory of Particle Astrophysics, Institute of High Energy Physics, Chinese Academy of Sciences, Beijing 100049, People's Republic of China}}

\correspondingauthor{\textbf{Qing-Cang Shui}}
\email{shuiqc@ihep.ac.cn}
\author[0000-0001-5160-3344]{\textbf{Qing-Cang Shui}}
\affiliation{\textbf{State Key Laboratory of Particle Astrophysics, Institute of High Energy Physics, Chinese Academy of Sciences, Beijing 100049, People's Republic of China}}
\affiliation{\textbf{University of Chinese Academy of Sciences, Chinese Academy of Sciences, 
100049 Beijing, People's Republic of China}}

\author[0000-0002-6454-9540]{\textbf{Peng-Ju Wang}}
\affiliation{\textbf{Institute f{\"u}r Astronomie und Astrophysik, Kepler Center for Astro and Particle Physics, 
Eberhard Karls Universit{\"a}t, Sand 1, D-72076 T{\"u}bingen, Germany}}

\author[0000-0001-9599-7285]{\textbf{Long Ji}}
\affiliation{\textbf{School of Physics and Astronomy, Sun Yat-Sen University, Zhuhai, 519082, People's Republic of China}}

\author[0000-0003-3188-9079]{\textbf{Ling-Da Kong}}
\affiliation{\textbf{Institute f{\"u}r Astronomie und Astrophysik, Kepler Center for Astro and Particle Physics, 
Eberhard Karls Universit{\"a}t, Sand 1, D-72076 T{\"u}bingen, Germany}}

\author[0000-0001-5586-1017]{\textbf{Shuang-Nan Zhang}}
\affiliation{\textbf{State Key Laboratory of Particle Astrophysics, Institute of High Energy Physics, Chinese Academy of Sciences, Beijing 100049, People's Republic of China}}
\affiliation{\textbf{University of Chinese Academy of Sciences, Chinese Academy of Sciences, 
100049 Beijing, People's Republic of China}}

\author[0000-0001-7584-6236]{\textbf{Hua Feng}}
\affiliation{\textbf{State Key Laboratory of Particle Astrophysics, Institute of High Energy Physics, Chinese Academy of Sciences, Beijing 100049, People's Republic of China}}

\author[0000-0001-8768-3294]{\textbf{Yu-Peng Chen}}
\affiliation{\textbf{State Key Laboratory of Particle Astrophysics, Institute of High Energy Physics, Chinese Academy of Sciences, Beijing 100049, People's Republic of China}}

\author[0000-0002-3776-4536]{\textbf{Ming-Yu Ge}}
\affiliation{\textbf{State Key Laboratory of Particle Astrophysics, Institute of High Energy Physics, Chinese Academy of Sciences, Beijing 100049, People's Republic of China}}
\affiliation{\textbf{University of Chinese Academy of Sciences, Chinese Academy of Sciences, 
100049 Beijing, People's Republic of China}}

\author[0000-0002-5554-1088]{\textbf{Jing-Qiang Peng}}
\affiliation{\textbf{State Key Laboratory of Particle Astrophysics, Institute of High Energy Physics, Chinese Academy of Sciences, Beijing 100049, People's Republic of China}}
\affiliation{\textbf{University of Chinese Academy of Sciences, Chinese Academy of Sciences, 
100049 Beijing, People's Republic of China}}

\author[0009-0001-3113-586X]{\textbf{Wen-zhong Li}}
\affiliation{\textbf{State Key Laboratory of Particle Astrophysics, Institute of High Energy Physics, Chinese Academy of Sciences, Beijing 100049, People's Republic of China}}
\affiliation{\textbf{University of Chinese Academy of Sciences, Chinese Academy of Sciences, 
100049 Beijing, People's Republic of China}}




\begin{abstract}
We have conducted a detailed spectral analysis of Swift J0243.6+6124 in its sub‑Eddington regime observed by Insight-HXMT and NICER during a series of outbursts including the giant one in 2018, and discovered a new transition at \(L_{\rm t}\sim4.5\times10^{37}\,\mathrm{erg\,s^{-1}}\)  accompanied by the evolution of the spectral parameters, in particular a significant turnover of the blackbody normalization. \(L_{\mathrm{t}}\) in the sub-Eddington regime represents the fifth transition luminosity identified so far, further increasing the complexity of Swift~J0243.6+6124 and may be accounted via introducing a multipolar magnetic field configuration, where weak (\(\sim2.8\times10^{12}\,\mathrm{G}\)) and strong (\(\sim1.6\times10^{13}\,\mathrm{G}\)) magnetic poles govern the emission at different accretion rates. Such a magnetic field configuration is equivalent to a relatively weak pure dipole magnetic field of \(\sim6.6\times10^{12}\,\mathrm{G}\) on the scale of the magnetosphere radius, and allows the local magnetic field of the neutron star to exceed \(10^{13}\,\mathrm{G}\).
\end{abstract}



\section{Introduction} \label{sec1}

Accreting X-ray pulsars (AXRPs) in high-mass X-ray binaries are highly magnetized neutron stars that accrete matter from their early-type companions through either stellar wind capture or Roche-lobe overflow \citep{2022arXiv220414185M}. 
The strong magnetic field, typically \(B \sim 10^{12}\text{--}10^{13}\,\mathrm{G}\), truncates the accretion flow at the magnetospheric radius \(R_{\rm m}\), where the infalling plasma is forced to follow the magnetic field lines and is directed toward the magnetic poles. There, the material releases its gravitational potential energy as intense X-ray radiation \citep{2019A&A...622A..61S}. Extensive theoretical studies have established that the geometry and physical conditions above the polar caps evolve markedly with the mass accretion rate, \(\dot{M}\), giving rise to complex variations in both timing and spectral properties during outbursts\citep{1975A&A....42..311B,1991ApJ...367..575B,2012A&A...544A.123B,2015MNRAS.447.1847M,2022ApJ...939...67B}. Such evolutions are generally associated with characteristic luminosities.

Transitional luminosities are ubiquitous in AXRPs, with diverse physical origins proposed in the literature\citep{2013A&A...551A...1R,2024ApJ...963...42X}. At \(L_{\rm{X}}\sim10^{37–38}\,\mathrm{erg\,s^{-1}}\), the so-called critical luminosity, \(L_{\mathrm{crit}}\), is often invoked to explain abrupt changes in the spectral continuum\citep{2012A&A...544A.123B}. Theoretically, once \(L_{\rm{X}}\) exceeds \(L_{\mathrm{crit}}\), the centroid energy of the cyclotron resonance scattering feature (CRSF), \(E_{\mathrm{cyc}}\), is predicted to switch from a positive to a negative correlation. This bimodality coincides with transitions in hardness–intensity diagrams (HIDs) and breaks in the evolution of spectral parameters\citep{2013A&A...551A...1R,2019A&A...622A..61S}.

In the subcritical regime (\(L_{\rm{X}}<10^{37}\,\mathrm{erg\,s^{-1}}\)), spectral evolution may be driven by multiple mechanisms. Far below \(L_{\mathrm{crit}}\) (e.g., \(L_{\rm{X}}\sim10^{34-35}\,\mathrm{erg\,s^{-1}}\)), radiation-dominated shock may collapse into collisionless shock and vacuum polarization effects may dominate, as discussed for the extremely low transition luminosity in 1A\,0535+262 \citep{1982ApJ...257..733L,2024ApJ...963...42X}. As the mass‐accretion rate increases to yield \(L_{\rm{X}}\sim10^{36}\text{–}10^{37}\,\mathrm{erg\,s^{-1}}\), Coulomb interaction become the primary deceleration mechanism after the gas shock, defining a Coulomb luminosity, \(L_{\mathrm{Coul}}\)\citep{2012A&A...544A.123B,1982ApJ...257..733L}. In this regime, matter decelerates above the polar cap via Coulomb interaction and emits thermal radiation from the surface hotspot; the hybrid emission pattern resulting from the coexistence of pencil and fan beams persists until \(L_{\rm crit}\).
\citep{1991ApJ...367..575B}. Overall, above \(L_{\rm{X}}\sim10^{36}\,\mathrm{erg\,s^{-1}}\), only \(L_{\mathrm{Coul}}\) and \(L_{\mathrm{crit}}\) serve as viable transition luminosities, while observationally most AXRPs exhibit only a single spectral transition\citep{2013A&A...551A...1R}.

The Swift J0243.6+6124 was discovered by the Swift/BAT telescope at a flux of 80~mCrab in October~2017 \citep{2017ATel10809....1K,2017GCN.21960....1C}. 
Subsequently, the giant outburst that occurred in 2017 reached a peak luminosity of \(L_{\rm X}\sim2\times10^{39}\,\mathrm{erg\,s^{-1}}\), calculated under the assumption of a distance of 6.8~kpc to the companion star, as reported by \citet{2018AJ....156...58B} based on Gaia DR2 parallax measurements. Even when adopting the latest Gaia DR3 distance of \(5.2 \pm 0.3\,\mathrm{kpc}\) (source ID:~465628193526364416; \citealt{2021AJ....161..147B}), the peak luminosity remains at \(L_{\rm{X}} \sim 1 \times 10^{39}\,\mathrm{erg\,s^{-1}}\), still exceeding the Eddington limit for a typical neutron star, confirming Swift J0243.6+6124 as the first Galactic ultraluminous X-ray pulsar \citep{2018MNRAS.479L.134T}. Timing analysis revealed a spin period of \(P_{\rm spin}\sim9.8\,\mathrm{s}\) and an optical counterpart classified as an O9.5 Ve star, confirming its BeXRB nature\citep{2017ATel10809....1K,2017ATel10866....1B,2017GCN.21960....1C,2018A&A...613A..19D,2017ATel10822....1K,2017ATel10815....1Y,2020A&A...640A..35R}. 

Since discovery, Swift J0243.6+6124 has undergone a series of subsequent outbursts. Even at the faintest luminosity of \(L_{\rm{X}}\sim3\times10^{34}\,\mathrm{erg\,s^{-1}}\)( \(0.1-200\,\mathrm{keV},\,6.8\,\mathrm{kpc}\)), NuSTAR detected no propeller effect, demonstrating that its accretion luminosity covers six orders of magnitude (\(10^{34}\text{–}10^{39}\,\mathrm{erg\,s^{-1}}\)), and making it an exemplary source for studying the full accretion evolution in XRPs\citep{2020MNRAS.491.1857D}.

In the sub‐Eddington regime, the broadband X‐ray spectrum of Swift~J0243.6+6124 can be well described by an absorbed cutoffpl and a blackbody component. As the luminosity increases, the Fe K$\alpha$ emission becomes prominent, exhibiting a narrow‐to‐broad line transition\citep{2019ApJ...885...18J,2020ApJ...902...18K,2017ATel10866....1B,2018MNRAS.474.4432J}; two additional low‐temperature blackbody components gradually emerge, which may correspond to radiation from an optically thick outflow and from the top of the accretion column\citep{2019ApJ...873...19T}. The optically thick outflow is likely produced by the transition from a gas‐pressure‐dominated (GPD) to a radiation‐pressure‐dominated (RPD) disc. \citep{2019ApJ...873...19T,2019MNRAS.487.4355V,2020MNRAS.491.1857D,2020MNRAS.497.5498W}.
\begin{figure}[!htbp]
    \centering
    \includegraphics[width=\linewidth]{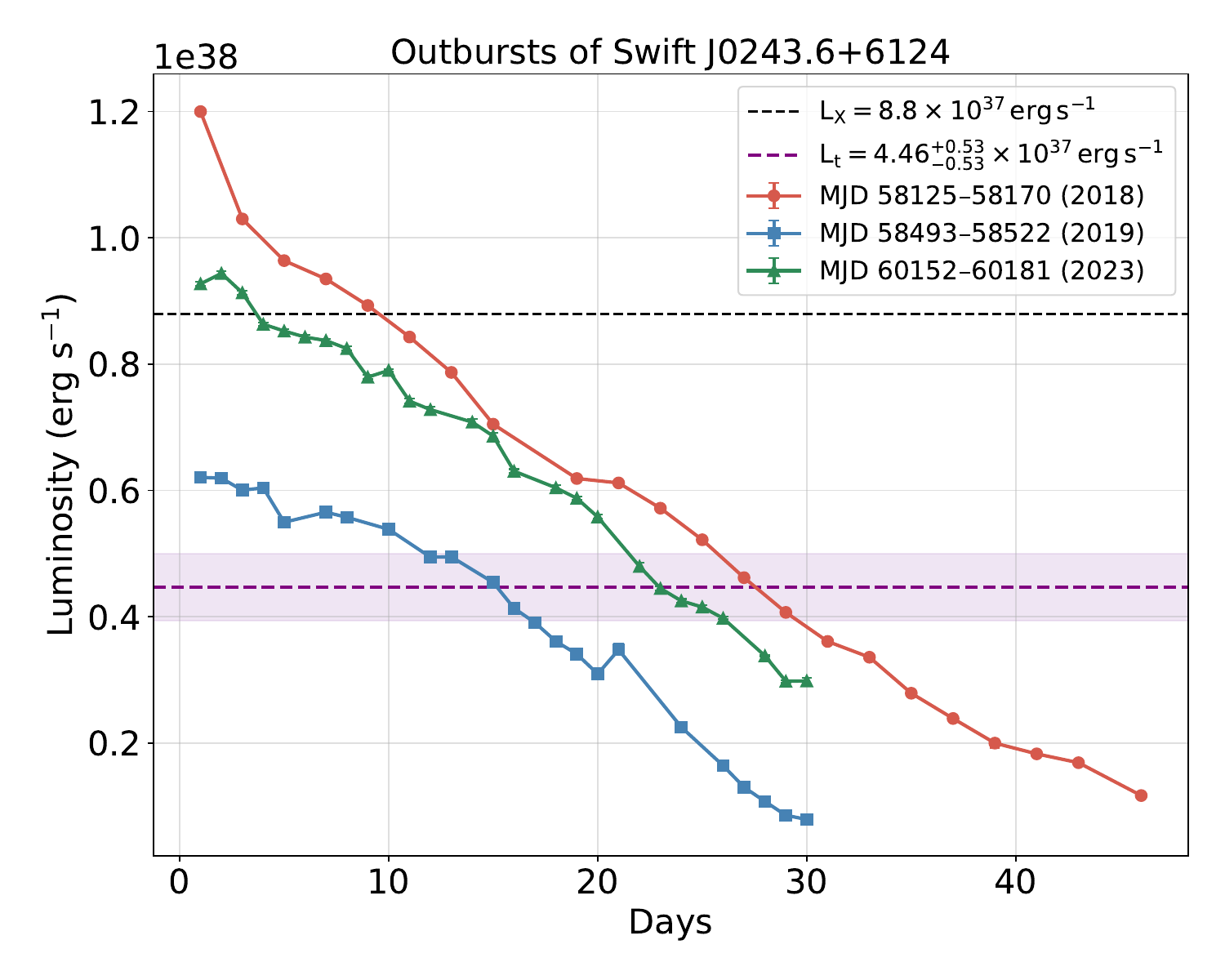}
    \caption{
        Bolometric luminosity evolution during the decay phases of three outbursts. The luminosity is derived from Insight-HXMT observations in the 2–150 keV band. The red, blue, and green curves correspond to the outbursts in 2018 (MJD 58125–58170), 2019 (MJD 58493–58523), and 2023 (MJD 60152–60181), respectively. The purple dashed line marks the newly identified transition luminosity(with the uncertainty interval at the 90\% confidence level marked by the purple bar), while the black dashed line represents the previously reported critical luminosity $L_{\rm 1}$ of the source\citep{2020ApJ...902...18K}. The x-axis shows the elapsed days since the first data point used for each outburst.
    }
    \label{fig1}
\end{figure}

Several transition luminosities have been discovered in Swift J0243.6+6124. From the pulse fraction evolution one infers a transition luminosity of \(\sim7.7\times10^{38}\,\mathrm{erg\,s^{-1}}\)( \(2-150\,\mathrm{keV},\,6.8\,\mathrm{kpc}\), using Insight-HXMT)\citep{2020MNRAS.497.5498W}. Insight‐HXMT observations of both spectral parameters and pulse‐profile evolution confirm two spectral transition luminosities at \(L_{\rm{1}}\sim1.5\times10^{38}\,\mathrm{erg\,s^{-1}}\) and \(L_{\rm{2}}\sim4.4\times10^{38}\,\mathrm{erg\,s^{-1}}\)( \(2-150\,\mathrm{keV},\,6.8\,\mathrm{kpc}\)); the latter is interpreted as the critical luminosity, yielding a magnetic field estimate of \(B\sim2.4\times10^{13}\,\mathrm{G}\) , which is  consistent with the discovery of 146keV CRSF line and at odds with weaker dipolar field estimates of a few \(\times10^{12}\,\mathrm{G}\), indicating a multipolar magnetic configuration\citep{2020ApJ...902...18K,2020MNRAS.491.1857D,2022ApJ...933L...3K,2018MNRAS.479L.134T}. Long‐term NICER monitoring also provides the full spectral parameter evolution and reveals transitions, but NICER’s narrowband yields results that differ markedly from broadband studies \citep{2020ApJ...902...18K,2024ApJ...963..132C}.At lower luminosities, the NICER pulse-profile evolution confirms a transition at \(L_{\rm t}\sim7\times10^{36}\,\mathrm{erg\,s^{-1}}\) (2–150~keV; 5.2~kpc), likely marking the onset of significant Coulomb deceleration\citep{2023MNRAS.522.6115S}.

\begin{figure*}[!htbp]
    \centering
    \begin{minipage}{0.28\textwidth}
        \centering
        \includegraphics[width=\textwidth]{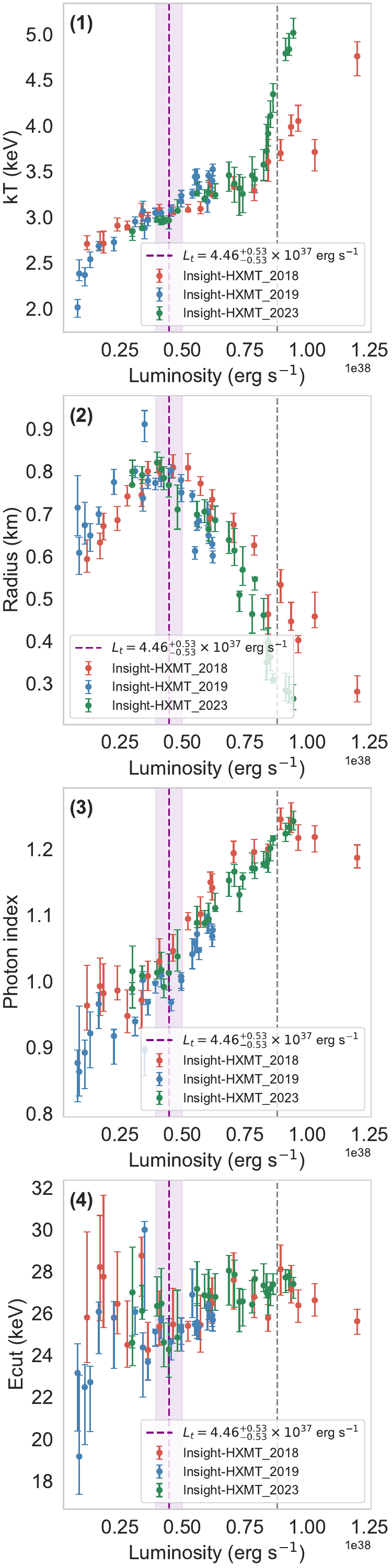}
        \vspace{0.2cm}  
        \makebox[0pt][c]{\textbf{(a) Insight-HXMT}}
    \end{minipage}
    \hspace{0.1in}
    \begin{minipage}{0.28\textwidth}
        \centering
        \includegraphics[width=\textwidth]{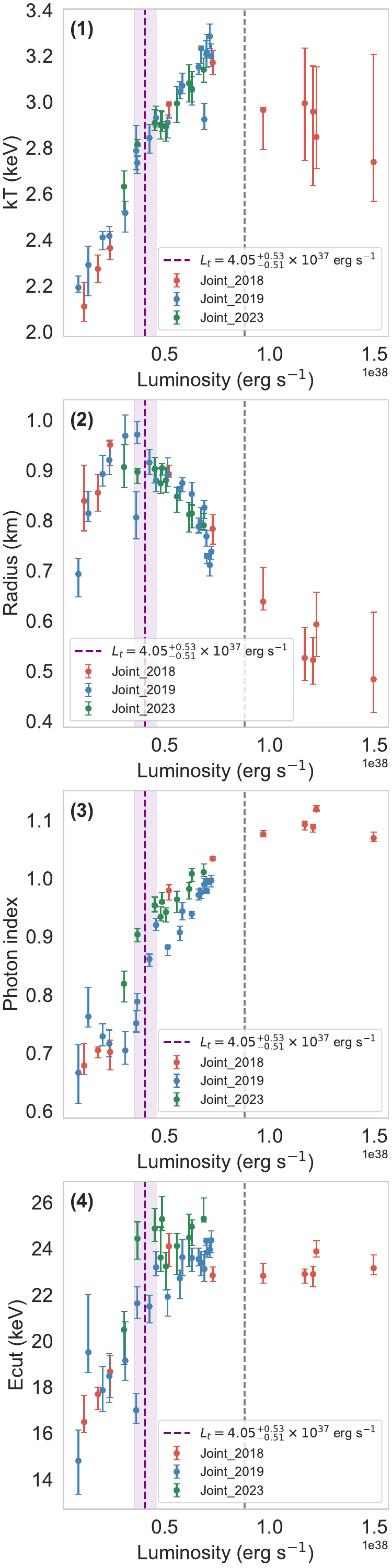}
        \vspace{0.2cm}  
        \makebox[0pt][c]{\textbf{(b) NICER + Insight-HXMT}}
    \end{minipage}
    \hspace{0.1in}
    \begin{minipage}{0.28\textwidth}
        \centering
        \includegraphics[width=\textwidth]{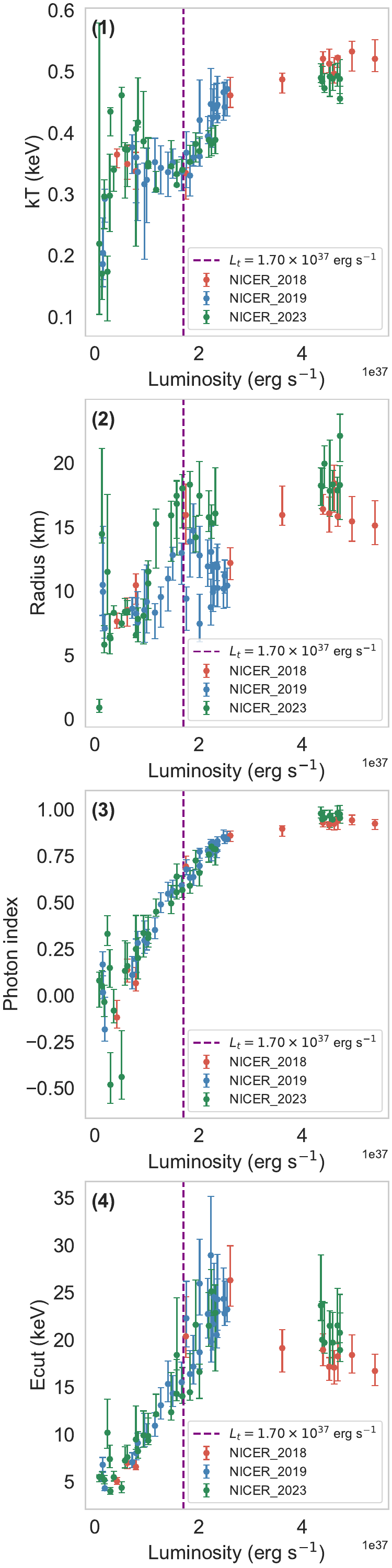}
        \vspace{0.2cm}  
        \makebox[0pt][c]{\textbf{(c) NICER}}
    \end{minipage}

    \vspace{0.2cm}  

    \caption{
    Parameter evolution versus luminosity derived from spectral fits performed with three instrument combinations.
    \textbf{Left:} Insight–HXMT only (LE: 2--10\,keV; ME: 8--30\,keV; HE: 28--100\,keV), covering 2--100\,keV in fitting, with luminosities evaluated over 2 -- 150\,keV;
    \textbf{Middle:} Joint fits in which NICER (0.7--10\,keV) replaces the LE data when observations are within 24\,hr (ME: 8--30\,keV; HE: 28--100\,keV); luminosities are also evaluated over 2 -- 150\,keV;
    \textbf{Right:} NICER-only narrowband fits (0.7--10\,keV), with luminosities computed in the same energy band. The grey dashed line marks the adopted transition luminosity $L_{\rm t}$ (0.7--10\,keV). The purple shaded band in the left and middle panels spans the 90\% credible range for each data set.
    }

    \label{fig2}
\end{figure*}

Between MJD\,58125 and MJD\,60181, one giant outburst and two normal outbursts of Swift~J0243.6+6124 were recorded by Insight–HXMT. In this paper, we present a detailed spectral analysis of Swift~J0243.6+6124—the first Galactic ultraluminous X-ray pulsar—in its sub-Eddington accretion regime, using sub-Eddington observations from Insight–HXMT complemented by contemporaneous NICER data. Considering that the updated distance affects the derived parameters throughout our analysis, we adopt the latest Gaia~DR3 distance of \(5.2 \pm 0.3\,\mathrm{kpc}\) \citep{2021AJ....161..147B} uniformly in this work. The rest of the paper is organized as follows: Section~\ref{sec2} describes the observations and data‐reduction strategy; Section~\ref{sec3} presents the luminosity and spectral‐evolution results; Section~\ref{sec4} discusses these findings; and Section~\ref{sec5} summarizes our conclusions.


\section{Observations and Data Reduction}
\label{sec2}

\subsection{Insight-HXMT}\label{sec2.1}
The Hard X-ray Modulation Telescope (Insight-HXMT) was launched on June 15, 2017. It features a wide energy band (1--250 keV), a large effective area at high energies, and is free from the pile-up effect for bright sources\citep{2014SPIE.9144E..21Z,2020SCPMA..6349502Z}. Insight-HXMT consists of three collimated telescopes: the High Energy X-ray telescope (HE), the Medium Energy X-ray telescope (ME), and the Low Energy X-ray telescope (LE). Their respective collecting areas and energy ranges are approximately \(5100\,\mathrm{cm}^2\) over 20--250\,keV for HE, \(952\,\mathrm{cm}^2\) over 5--30\,keV for ME, and \(384\,\mathrm{cm}^2\) over 1--10\,keV for LE.
The primary fields of view (FoV) of LE, ME, and HE are $1.6^\circ\times6^\circ$, $1^\circ\times4^\circ$, and $1.1^\circ\times5.7^\circ$, respectively \citep{2020SCPMA..6349503L,2020SCPMA..6349504C,2020SCPMA..6349505C}.

From October 7, 2017 (MJD 58033) to February 21, 2018 (MJD 58170), Insight-HXMT triggered a total of 122 pointed exposures, sampling the entire giant outburst of Swift J0243.6+6124. In subsequent outbursts, Insight-HXMT observed the decay phases of two outbursts in 2019 and 2023, covering January 10, 2019 (MJD 58493) to February 9, 2019 (MJD 58523) and July 27, 2023 (MJD 60152) to August 25, 2023 (MJD 60181), respectively. Since the 2019 and 2023 outbursts did not exceed the Eddington luminosity (for a typical neutron star, \(L_{\rm Edd}\sim1.8\times10^{38}\,\mathrm{erg\,s^{-1}}\) in the 2 -- 150\,keV band) and our focus is on the sub-Eddington regime, we selected observations from three decay intervals: January 7, 2018 (MJD 58125) to February 21, 2018 (MJD 58170), January 10, 2019 (MJD 58493) to February 9, 2019 (MJD 58523), and July 27, 2023 (MJD 60152) to August 25, 2023 (MJD 60181).
The bolometric luminosity spans from $7.9\times10^{36}\;\mathrm{erg\,s}^{-1}$ to $1.2\times10^{38}\;\mathrm{erg\,s}^{-1}$(\(2-150\,\mathrm{keV},\,5.2\,\mathrm{kpc}\), see Figure~\ref{fig1}). The data in the luminosity range of \(1.2\times10^{38}\) – \(1.8\times10^{38}\,\mathrm{erg\,s^{-1}}\) have already been investigated in detail by \citet{2020ApJ...902...18K}, and are not discussed there.

In this work, we processed the data using the Insight-HXMT Data Analysis Software (HXMTDAS) v2.06 and applied barycenter correction and binary orbital correction using the \texttt{hxbary} and \texttt{binCor} tools, respectively. The data were filtered using the Good Time Intervals (GTIs) recommended by the Insight-HXMT team, and only data with at least 300 s before and after the South Atlantic Anomaly (SAA) passages were used. In addition, since our study focuses on the low-luminosity regime, to improve the fitting quality we combined the observational data on a daily basis and re-binned the spectra using \texttt{grppha}. Depending on the instrument performance and specific requirements, we selectively employed certain energy bands; details are provided in subsequent sections.

\subsection{NICER}\label{sec2.2}
The Neutron star Interior Composition Explorer (NICER) was launched in June 2017 and installed on the International Space Station. It is equipped with the X-ray Timing Instrument (XTI), designed to operate in the 0.2--12 keV energy range\citep{2016SPIE.9905E..1HG,2016SPIE.9905E..1IP}. NICER observed the giant and subsequent normal X-ray outbursts between MJD 58029 (October 3, 2017) and MJD 58533 (February 19, 2019) with a net exposure time of 408 ks. In addition, NICER observed the recent 2023 outburst, accumulating approximately 104 ks of net exposure from observations conducted between June and September 2023 under observation IDs 6050390227--6050390277. We selected NICER data overlapping with the Insight-HXMT observation periods for spectral fitting. The unfiltered event data from NICER were processed using the \texttt{nicerl2} pipeline within HEASOFT v6.30. The analysis was carried out with the gain version \texttt{xti20221001} under the condition that the calibration database files were available. Spectra were extracted using \texttt{nicerl3-spect} pipelines.

\begin{figure*}[!htbp]
    \centering
    \begin{minipage}{0.45\textwidth}
        \centering
        \includegraphics[width=\textwidth]{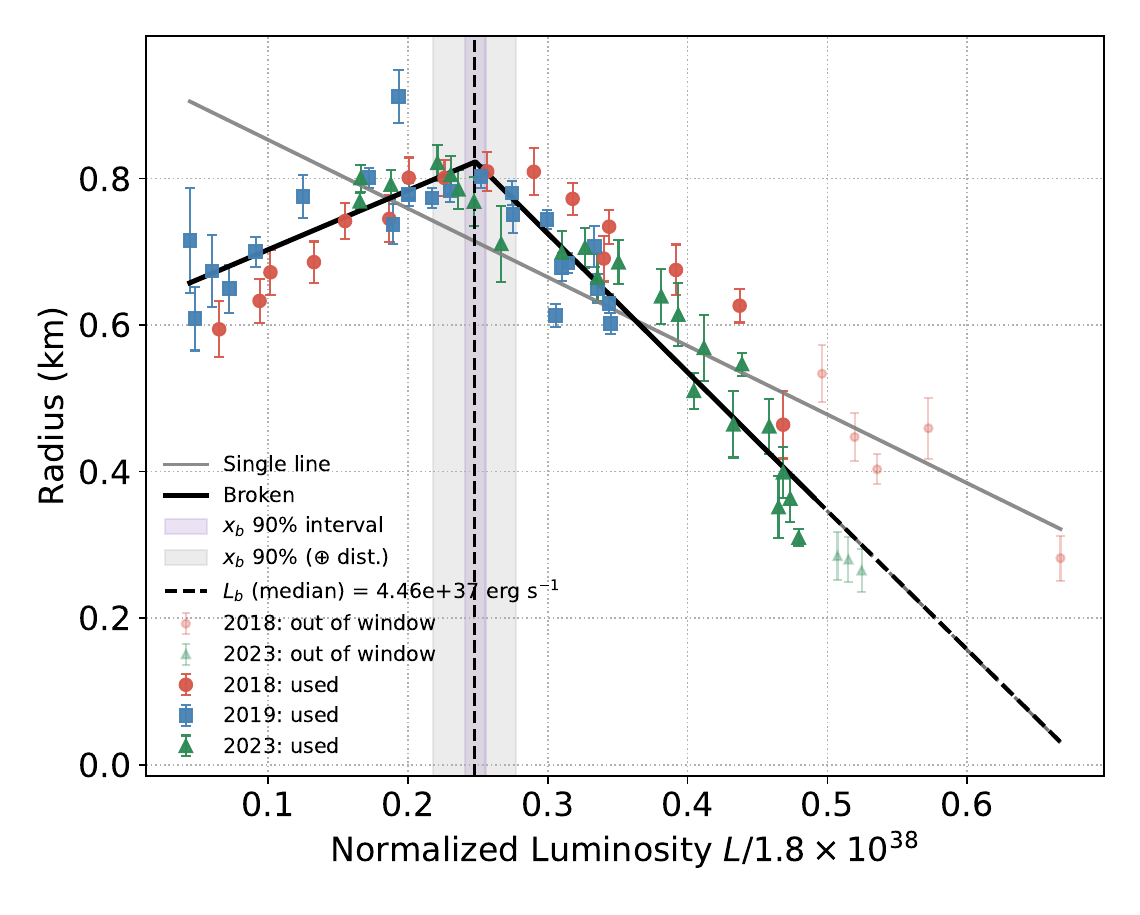}
        \vspace{0.2cm}  
    \end{minipage}
    \hspace{0.1in}
    \begin{minipage}{0.45\textwidth}
        \centering
        \includegraphics[width=\textwidth]{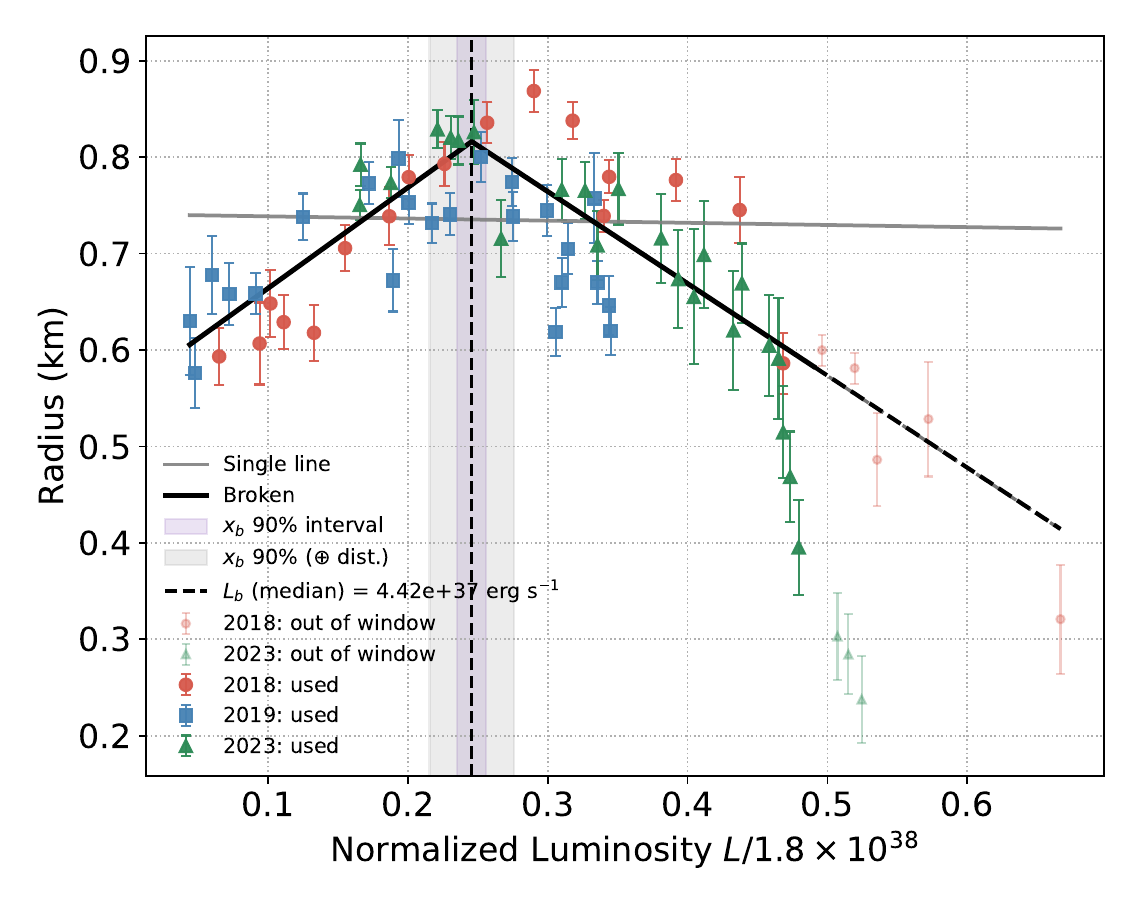}
        \vspace{0.2cm}  
    \end{minipage}
    \hspace{0.1in}

    \vspace{0.2cm}  

    \caption{\textbf{
    Broken--linear fit to $Radius$ versus normalized luminosity $x = L / (1.8 \times 10^{38})$ with $x \le 0.49$.
    Data points are color--coded by year; the solid curve represents the maximum--a--posteriori (MAP) broken--line model,
    and the vertical dashed line marks the break location.
    The gray and purple shaded bands show, respectively, the 90\% credible intervals for $x_b$  with and without the inclusion of distance uncertainty.
    \textit{Left panel} (Insight--HXMT, frozen $N_{\rm H}$): the break corresponds to
    $L_x = (4.46^{+0.53}_{-0.53}) \times 10^{37}\ {\rm erg\ s^{-1}}$.
    \textit{Right panel} (Insight--HXMT, thawed $N_{\rm H}$): the break is
    $L_x = (4.42^{+0.54}_{-0.54}) \times 10^{37}\ {\rm erg\ s^{-1}}$.}
    }

    \label{fig3}
\end{figure*}

\begin{figure*}[!htbp]
    \centering
    \begin{minipage}{0.45\textwidth}
        \centering
        \includegraphics[width=\textwidth]{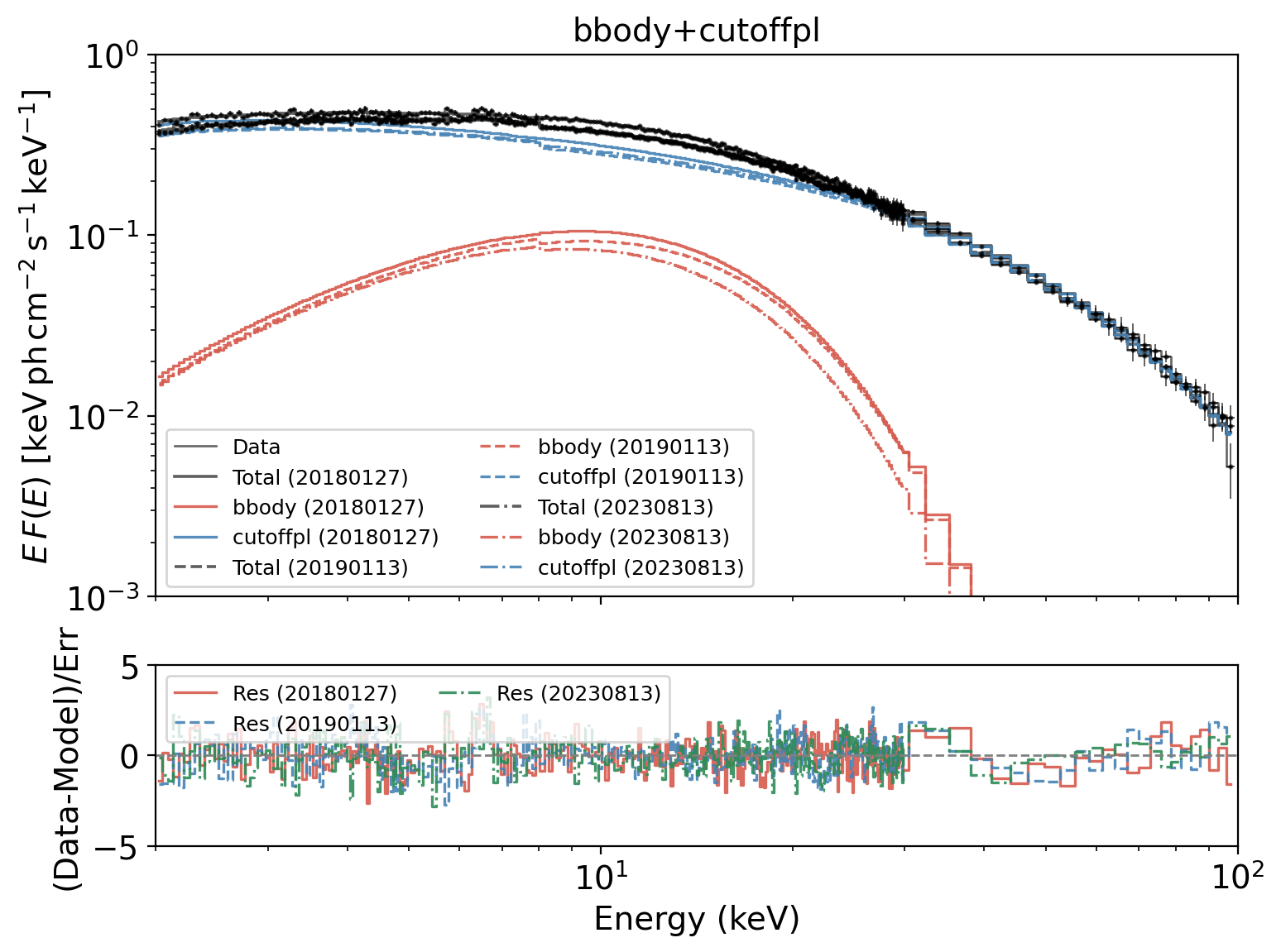}
        {\footnotesize\textbf{(a)}} 
    \end{minipage}
    \hspace{0.1in}
    \begin{minipage}{0.45\textwidth}
        \centering
        \includegraphics[width=\textwidth]{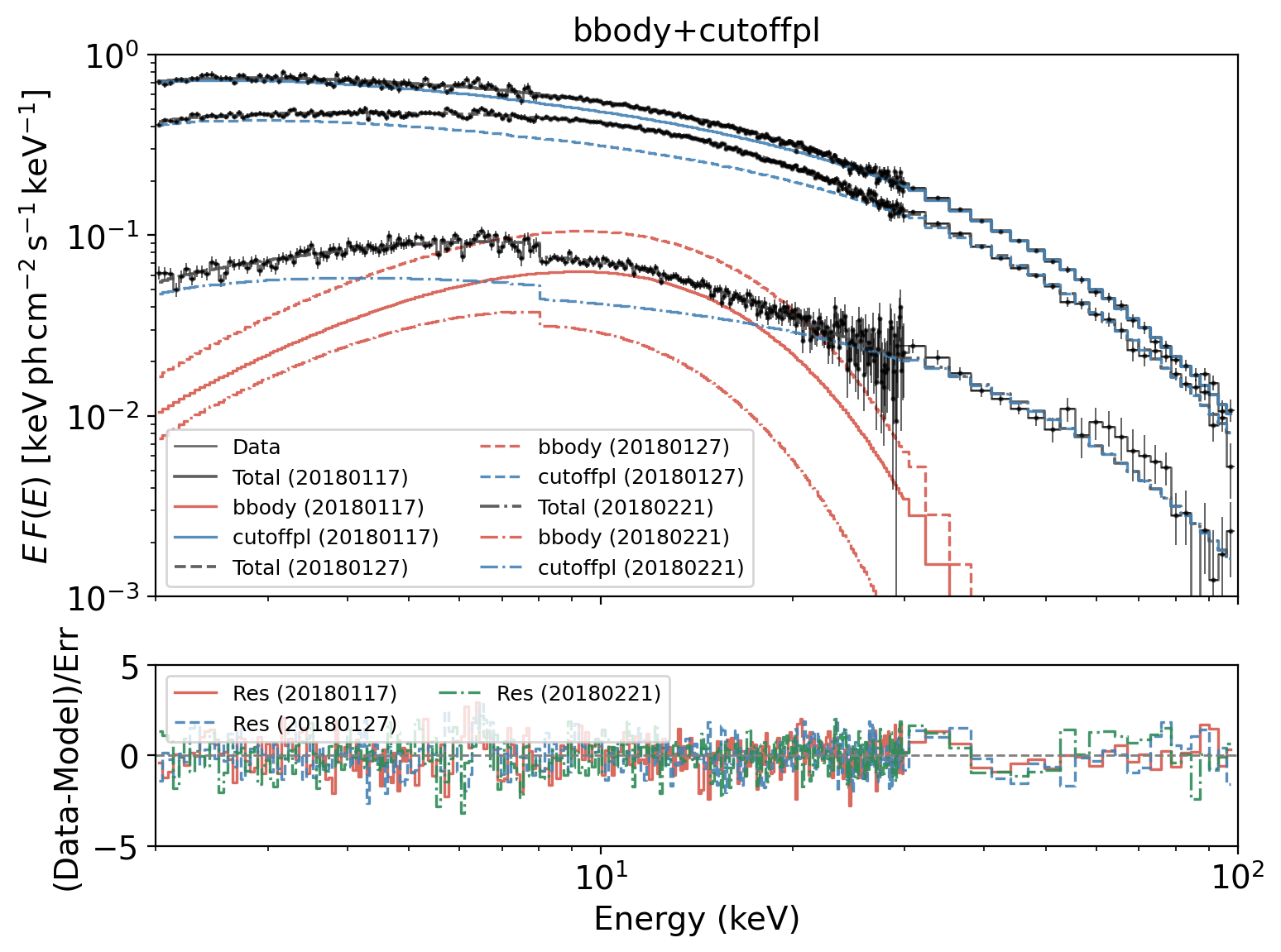}
        {\footnotesize\textbf{(b)}} 
    \end{minipage}
    \hspace{0.1in}
    \begin{minipage}{0.45\textwidth}
        \centering
        \includegraphics[width=\textwidth]{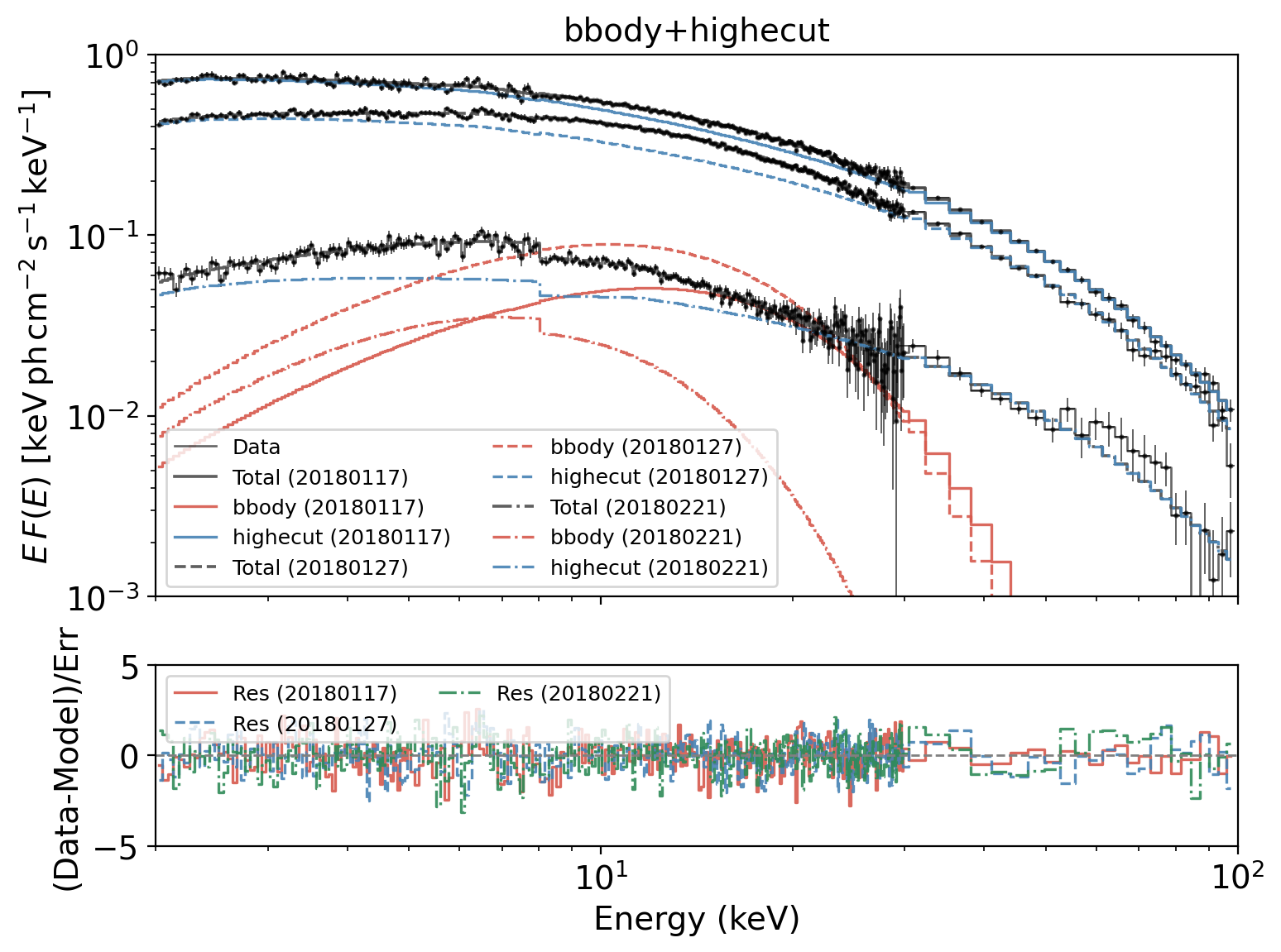}
        \vspace{0.2cm}  
        {\footnotesize\textbf{(c)}} 
    \end{minipage}
    \hspace{0.1in}
    \begin{minipage}{0.45\textwidth}
        \centering
        \includegraphics[width=\textwidth]{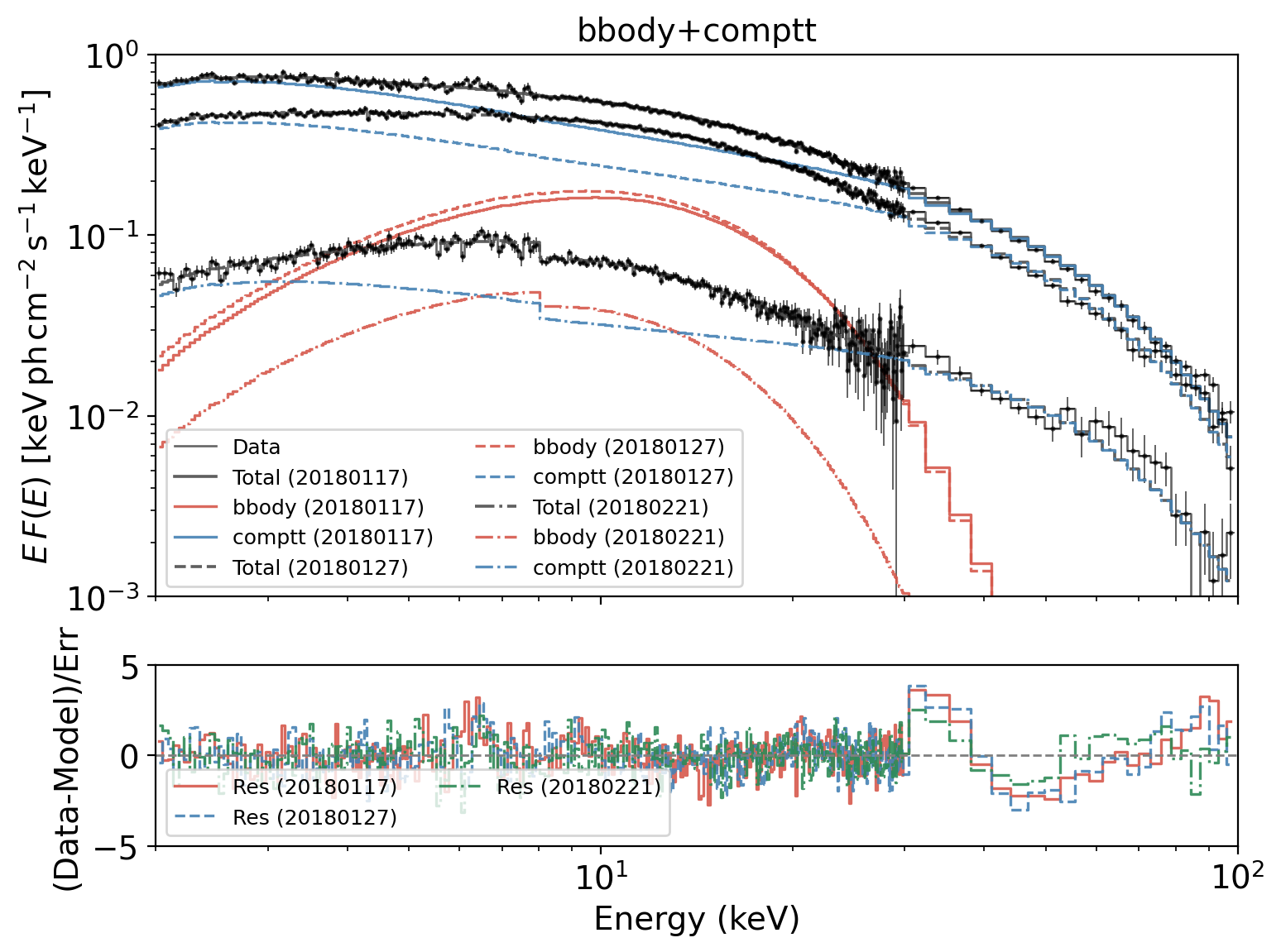}
        \vspace{0.2cm}  
        {\footnotesize\textbf{(d)}} 
    \end{minipage}
    \hspace{0.1in}
    \begin{minipage}{0.45\textwidth}
        \centering
        \includegraphics[width=\textwidth]{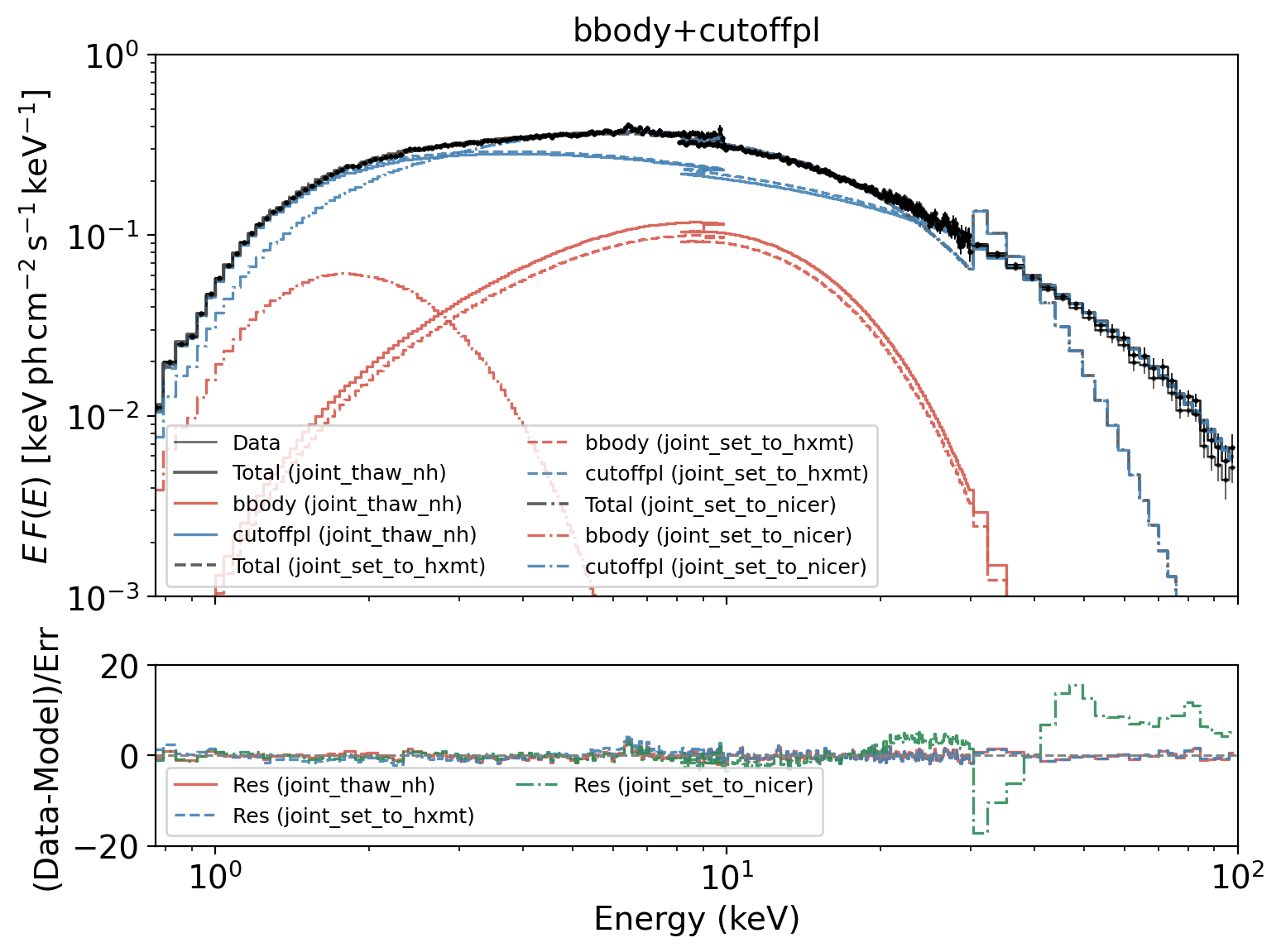}
        \vspace{0.2cm}  
        {\footnotesize\textbf{(e)}} 
    \end{minipage}
    \hspace{0.1in}
    \begin{minipage}{0.45\textwidth}
        \centering
        \includegraphics[width=\textwidth]{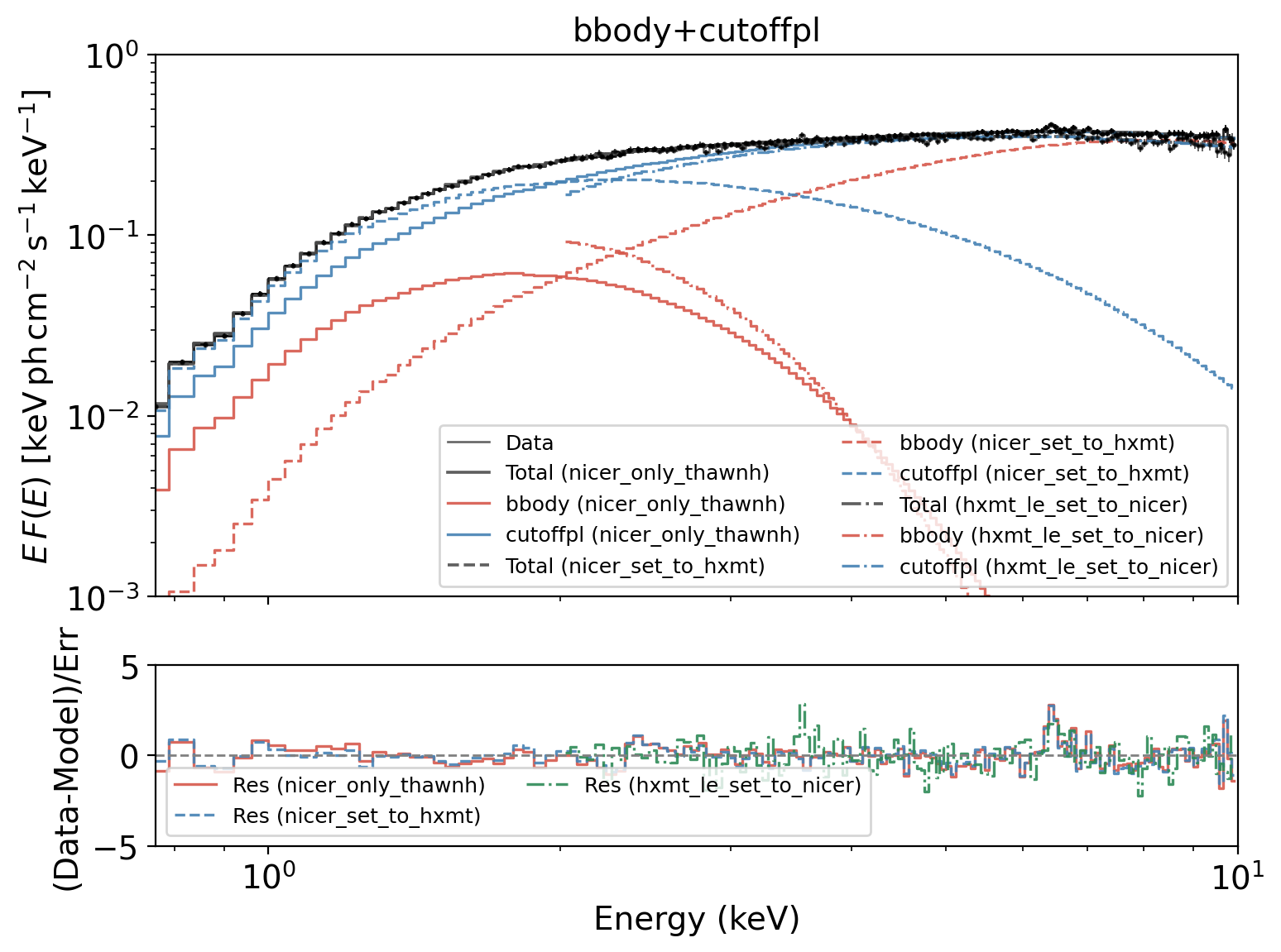}
        \vspace{0.2cm}  
        {\footnotesize\textbf{(f)}} 
    \end{minipage}
    \hspace{0.1in}

    \vspace{0.2cm}  

\caption{%
\textit{Panel (a):} Comparison of spectral fits at similar luminosities from three
observations in different outbursts, using Insight--HXMT data only.
\textit{Panels (b)–(d):} Spectral fits for three representative observations at
luminosities below, around, and above $L_{\rm t}$, respectively, obtained with
different continuum models; the corresponding best-fit parameters are listed in
Table~2.
\textit{Panel (e):} Joint spectral fitting of the 2018 February 02 observation
over 0.7--100~keV with $N_{\rm H}$ free, comparing three cases: (1) a fully free
joint fit; (2) a joint fit with the continuum fixed to the
Insight--HXMT-only best fit in 2--100~keV; and (3) a joint fit with the
continuum fixed to the NICER-only best fit in 0.7--10~keV. In cases (2) and (3),
$N_{\rm H}$, the normalization, and the cross-calibration constant are allowed to
vary.
\textit{Panel (f):} Fits to the 2018 February 02 NICER and Insight--HXMT
LE data with $N_{\rm H}$ free, comparing: (1) a free NICER-only fit in 0.7--10~keV;
(2) a free Insight--HXMT LE-only fit in 2--10~keV with parameters
initialized at the NICER best-fit values; and (3) a NICER-only fit initialized
at the best-fit parameters from the fully free joint fit.%
}

    \label{fig4}
\end{figure*}

\section{Results} \label{sec3}
\subsection{Luminosity Evolution}\label{sec3.1}

The spectrum of Swift~J0243.6+6124 was modeled using
\texttt{constant*tbabs*(bbodyrad+cutoffpl)} in XSPEC (version~12.13.1; \citealt{1996ASPC..101...17A}), where the multiplicative \texttt{constant} accounts for cross-calibration uncertainties among different instruments. Within the luminosity range explored in this work, the iron K$\alpha$ line near 6.4~keV cannot be well constrained and has a negligible impact on the best-fit results; therefore, no additional Gaussian component was included. Parameter uncertainties were estimated using Markov Chain Monte Carlo (MCMC) sampling with 20{,}000 iterations, 40 walkers, and a burn-in phase of
2{,}000 steps, and are quoted at the 90\% confidence level. We apply \texttt{constant*tbabs*cflux*(bbodyrad+cutoffpl)} to estimate the 2–150\,keV luminosity, assuming a distance of 5.2\,kpc reported in \citet{2021AJ....161..147B} based on Gaia DR3 parallax measurements of the companion star.

As can be seen, the luminosity ranges of the three results shown in Figure~\ref{fig2} are not identical. Here we provide a detailed description of the luminosity calculation adopted in this work. In Figure~\ref{fig2}(a), where only Insight--HXMT data in the 2--100~keV band are fitted, we use \texttt{cflux} to compute the source flux in the 2--150~keV band and convert it into luminosity. In Figure~\ref{fig2}(b), where a joint fit is performed over the 0.7--100~keV band, we calculate the source luminosity in the 2--150~keV band. In Figure~\ref{fig2}(c), where only NICER data in the 0.7--10~keV band are fitted, we compute the 0.7--10~keV luminosity, in order to reproduce the results reported by \citet{2024ApJ...963..132C}.

To maintain consistency with previous studies\citep{2020ApJ...902...18K,2020MNRAS.497.5498W,2022ApJ...933L...3K}, we primarily adopt the luminosity range obtained in Figure~\ref{fig2}(a) for our discussion. The luminosity range derived in Figure~\ref{fig2}(a) is \((7.9\text{--}120)\times10^{36}\,\mathrm{erg\,s^{-1}}\), which covers the decay phases of all three outbursts, and this range is shown in Figure~\ref{fig1}. We found that, in the 2--150\,keV energy band, the luminosities derived from the joint fit are systematically higher by about 10\% compared to the Insight--HXMT--only results. This offset is likely caused by the inclusion of the 0.7--2\,keV band in the joint fit and does not affect the main results of this work.

\begin{deluxetable*}{llccccccc}
\tabletypesize{\scriptsize}
\tablewidth{0pt}
\tablecaption{Model comparison and break luminosity results for all datasets.\label{tab:model_comparison}}
\tablehead{
\colhead{Group} & \colhead{Dataset} & \colhead{$\Delta\ln\mathcal{L}$} &
\colhead{$\Delta\mathrm{AICc}$} & \colhead{$\Delta\mathrm{BIC}$} &
\colhead{$p_{\mathrm{LRT}}$} &
\colhead{$L_{\mathrm{b,\,MAP}}$} &
\colhead{$L_{\mathrm{b,\,90\%\,CI}}$} &
\colhead{$\Delta L$} \\
\colhead{} & \colhead{} & \colhead{} & \colhead{} & \colhead{} & \colhead{} &
\colhead{($10^{37}\,\mathrm{erg\,s^{-1}}$)} &
\colhead{($10^{37}\,\mathrm{erg\,s^{-1}}$)} &
\colhead{($10^{37}\,\mathrm{erg\,s^{-1}}$)} 
}
\startdata
Group\,1 & Insight--HXMT                & 364.49 & 724.48 & 720.72 & $<10^{-4}$ & 4.46 & [3.93, 4.99] & $+0.00$ \\
Group\,2 & Insight--HXMT (Thaw--NH)     & 130.08 & 255.67 & 251.88 & $<10^{-4}$ & 4.42 & [3.88, 4.96] & $-0.04$ \\
Group\,3 & Joint                        & 77.21  & 149.47  & 147.32  & $<10^{-4}$ & 4.17 & [3.91, 4.82] & $-0.29$ \\
Group\,4 & Joint (cut\,0.7--2\,keV)     & 33.40  & 61.85  & 59.70  & $<10^{-4}$ & 5.33 & [4.63, 6.07] & $+0.87$ \\
\enddata
\tablecomments{All tests favor the broken--line model ($M_1$) over the single--line model ($M_0$). The $p$--values from the LRT (Likelihood--Ratio Test) are reported as upper limits, since none of the simulations exceeded the observed $\Delta\ln\mathcal{L}$, indicating that the improvement is statistically significant. Overall, the break luminosities derived from the four datasets are consistent within the uncertainties. The column $\Delta L$ is computed relative to the best-fit break luminosity of Group~1. Note that for Group~4, the spectral parameters were obtained by excluding the 0.7--2\,keV band, while the luminosities were still calculated following Case~2. 
As a result, the luminosities are systematically higher by about 10\% compared to those derived in Case~1 (see Section~3.1). If a 10\% systematic uncertainty is taken into account for Groups~3 and~4, the corresponding break luminosities become $L_{\mathrm{b}} = 3.75\,[3.52,\,4.34]$ and $4.80\,[4.17,\,5.46]$, respectively, 
which are consistent with those of Groups~1 and~2 within the uncertainties.}
\label{table1}
\end{deluxetable*}

\subsection{Spectral Discrepancy with Different Telescope Configurations}\label{sec3.2}
\begin{figure}[!htbp]
    \centering
    \includegraphics[width=\linewidth]{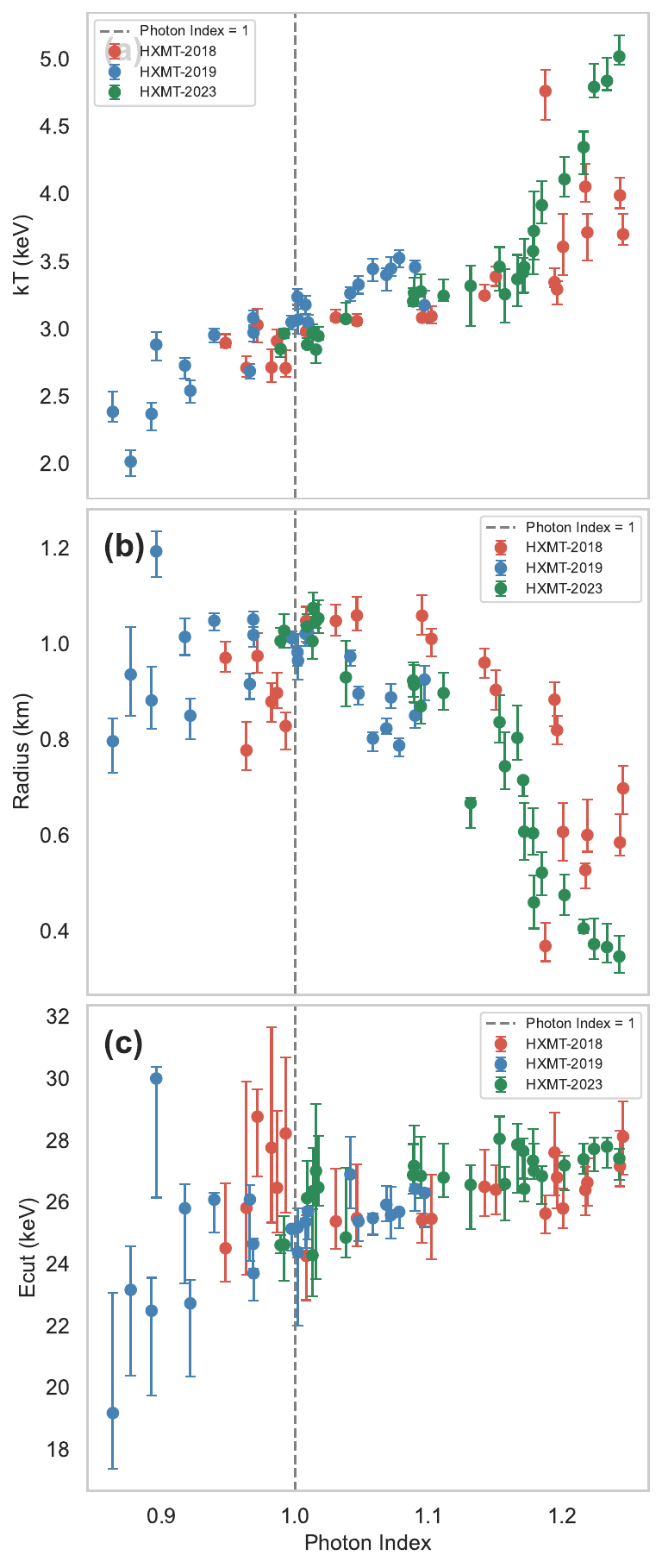}
    \caption{
       Correlation between model parameters and the photon index. The grey dashed line represents Photon Index = 1.0, corresponding to $L_t$.
    \label{fig5}
    }
\end{figure}
\citet{2024ApJ...963..132C} performed spectral fitting of long‐term NICER data in the 0.7–10\,keV band for Swift~J0243.6+6124, deriving spectral‐parameter evolutions over a luminosity range similar to that observed by Insight–HXMT\citep{2020ApJ...902...18K}, yet with markedly different values and trends. To address this discrepancy, we compare the spectral evolutions with data of NICER only, Insight-HXMT only and the joint Insight-HXMT /NICER observations, with results shown in Figure~\ref{fig2}.

For Insight–HXMT we restrict the analysis to the 2–100\,keV band (LE: 2–10\,keV; ME: 8–30\,keV; HE: 28–100\,keV), omitting \(>100\)\,keV data dominated by background and 1–2\,keV data showing instrumental  artifacts. For NICER we adopt the 0.7–10\,keV band to avoid calibration uncertainties below 0.4\,keV, above 10.0\,keV, and the edge‐like feature near 0.5\,keV\citep{2024ApJ...963..132C}. Absorption is modeled with the Wilms abundance table and Verner photoionization cross sections and fits are evaluated using \(\chi^2\) statistics\citep{2000ApJ...542..914W}. Based on Insight–HXMT calibration and previous NICER work, spectra are rebinned as follows: (1) NICER: a minimum of 30 counts per energy bin; (2) LE: channels 0–579 grouped in 5, channels 580–1535 in 10; (3) ME: channels 0–1023 grouped in 2; (4) HE: channels 0–255 grouped in 2. A \texttt{constant} component accounts for inter‐instrument calibration, and the final model is \texttt{constant*tbabs*(bbodyrad+cutoffpl)}.

The parameter settings for each detector combination are as follows:

1. Insight--HXMT only: we use the 2--100\,keV band with a systematic error of 1\%. We initially allowed \(n_{\rm H}\) to vary and found a mean value of \(\sim 1.0\times10^{22}\,\mathrm{cm^{-2}}\), consistent with \citet{2024ApJ...963..132C}. For consistency, we fixed \(n_{\rm H}=1.064\times10^{22}\,\mathrm{cm^{-2}}\) in all fits. The LE normalization constant was fixed at unity, while the ME and HE constants were left free to vary.

2.   NICER only: we use the 0.7–10\,keV band with a systematic error of 1.5\%  and fix \(n_{\rm H}=1.064\times10^{22}\,\mathrm{cm^{-2}}\) to reproduce previous analyses in \citet{2024ApJ...963..132C}.

3. Joint NICER+Insight–HXMT: we extend the bandpass to 0.7–100\,keV, fix the NICER constant at unity and set \(n_{\rm H}=1.064\times10^{22}\,\mathrm{cm^{-2}}\).

The spectral-evolution results for the three telescope configurations are shown in Figure~\ref{fig2}. In Group (a) (Insight–HXMT only) the reduced-\(\chi^2\) remains \(\sim1\) throughout the full luminosity range, indicating that the model adequately captures the spectral shape. In Group (b) (joint fits), the trends are consistent with those in Group (a), but for \(L_{\rm X} \gtrsim 8.8\times10^{37}\,\mathrm{erg\,s^{-1}}\), the reduced \(\chi^2\) increases from about 1.0 to 1.5. We find that adding an additional Gaussian iron-line component can reduce the reduced \(\chi^2\), but this part does not affect the main results of this paper and is beyond the scope of discussion, so we do not investigate it further.

The first four panels of Figure~\ref{fig4} display the fitting results for the representative Group (a) data. The (a) panel of Figure~\ref{fig4} shows that the spectra of different outbursts observed with Insight-HXMT can be well described by our model, and no significant differences are found at the same luminosity. Furthermore, We find that group (c), the narrowband fits (NICER only), which are consistent with those reported in \citet{2024ApJ...963..132C}, cannot constrain the spectral parameters very well. In group (c), all parameter values and their evolutionary trends diverge markedly, the \texttt{bbodyrad} component drifts to a low‐temperature, large‐scale solution (\(kT\sim0.4\ \mathrm{keV}\), \(R\sim20\ \mathrm{km}\)), while the other groups converge on a high‐temperature, small‐scale blackbody (\(kT\sim3\ \mathrm{keV}\), \(R\sim1\ \mathrm{km}\)).

We performed a restricted-fitting test on the 2018 February 02 observation. Prior to the joint fit over the 0.7--100~keV band, the continuum parameters were separately fixed to the best-fit values obtained from the
Insight--HXMT (2--100~keV) and NICER (0.7--10~keV) single-instrument fits (see Figure~\ref{fig4}(e); the free-fit results for the three instrumental configurations are summarized in Table~\ref{table3}). Fixing the continuum to the Insight--HXMT best-fit values yields results consistent with the fully free joint fit, whereas fixing it to the NICER best-fit values produces pronounced residual structures above $\sim$20~keV, indicating that the spectral shape inferred from NICER is incompatible with the high-energy data.

We further examine the NICER results (see Figure~\ref{fig4}(f)) and find that the continuum shape obtained from a free fit to the Insight--HXMT LE data in the 2--10~keV band is consistent with the NICER spectrum. However, when the initial parameter values of the NICER fit are set to those from the joint-fit, the resulting fit is clearly inconsistent with both the joint-fit and NICER-only results, indicating that model degeneracy alone is insufficient to account for the spectral bias introduced by narrow-band fitting. We therefore conclude that fitting with NICER data alone can intrinsically bias the inferred continuum and lead to
nonphysical results.


In summary, although the spectral results of \citet{2024ApJ...963..132C} provide supporting evidence for the outburst evolution of Swift~J0243.6+6124, the parameters are not well constrained because of the limited bandwidth. We base our subsequent analysis primarily on the Insight–HXMT and joint‐fit results.

\subsection{Spectral Evolution in Sub-Eddington Regime}\label{sec3.3}


\begin{deluxetable*}{llccccccccc}
\tablecaption{Spectral Parameters for Three Observations\label{tab:spec_three_obs}}
\tabletypesize{\footnotesize}           
\tablehead{
\colhead{Component} & \colhead{Parameter} &
\multicolumn{3}{c}{20180221 (\(\sim1.17\times10^{37}\,\mathrm{erg\,s^{-1}}\))} &
\multicolumn{3}{c}{20180127 (\(\sim6.12\times10^{37}\,\mathrm{erg\,s^{-1}}\))} &
\multicolumn{3}{c}{20180117 (\(\sim8.43\times10^{37}\,\mathrm{erg\,s^{-1}}\))} \\
\colhead{} & \colhead{} &
\colhead{Model 1} & \colhead{Model 2} & \colhead{Model 3} &
\colhead{Model 1} & \colhead{Model 2} & \colhead{Model 3} &
\colhead{Model 1} & \colhead{Model 2} & \colhead{Model 3}
}
\startdata
Tbabs    & $N_{\rm H}$ ($10^{22}\,{\rm cm^{-2}}$) &
  1.08$^{+0.09}_{-0.04}$ & 0.92$^{+0.20}_{-0.19}$ & 2.02$^{+0.04}_{-0.04}$ &
  0.73$^{+0.04}_{-0.04}$ & 0.69$^{+0.20}_{-0.24}$ & 1.64$^{+0.01}_{-0.01}$ &
  0.63$^{+0.14}_{-0.02}$ & 0.59$^{+0.24}_{-0.14}$ & 1.81$^{+0.25}_{-0.09}$ \\
bbodyrad & $kT$ (keV) &
  2.71$^{+0.14}_{-0.06}$ & 2.75$^{+0.12}_{-0.12}$ & 2.89$^{+0.05}_{-0.06}$ &
  3.29$^{+0.08}_{-0.08}$ & 3.28$^{+0.06}_{-0.03}$ & 3.20$^{+0.03}_{-0.03}$ &
  3.24$^{+0.25}_{-0.22}$ & 3.24$^{+0.06}_{-0.23}$ & 3.41$^{+0.04}_{-0.03}$ \\
bbodyrad & norm &
  1.30$^{+0.06}_{-0.19}$ & 1.18$^{+0.08}_{-0.20}$ & 1.41$^{+0.02}_{-0.02}$ &
  2.02$^{+0.12}_{-0.06}$ & 2.03$^{+0.12}_{-0.16}$ & 3.60$^{+0.11}_{-0.07}$ &
  1.27$^{+0.11}_{-0.16}$ & 1.28$^{+0.25}_{-0.11}$ & 2.85$^{+0.21}_{-0.32}$ \\
cutoffpl & $\Gamma$ &
  0.97$^{+0.08}_{-0.04}$ & \nodata & \nodata &
  1.10$^{+0.02}_{-0.01}$ & \nodata & \nodata &
  1.14$^{+0.03}_{-0.01}$ & \nodata & \nodata \\
cutoffpl & $E_{\rm cut}$ (keV) &
  25.9$^{+2.6}_{-0.7}$ & \nodata & \nodata &
  25.3$^{+1.0}_{-0.3}$ & \nodata & \nodata &
  24.7$^{+0.8}_{-0.3}$ & \nodata & \nodata \\
cutoffpl & norm &
  0.07$^{+0.01}_{-0.01}$ & \nodata & \nodata &
  0.58$^{+0.02}_{-0.01}$ & \nodata & \nodata &
  1.00$^{+0.04}_{-0.01}$ & \nodata & \nodata \\
highecut & $E_{\rm break}$ (keV) &
  \nodata & 4.8$^{+2.3}_{-1.0}$ & \nodata &
  \nodata & 2.3$^{+0.3}_{-0.7}$ & \nodata &
  \nodata & 2.4$^{+0.3}_{-0.8}$ & \nodata \\
highecut & $E_{\rm fold}$ (keV) &
  \nodata & 26.5$^{+1.0}_{-2.1}$ & \nodata &
  \nodata & 25.3$^{+0.9}_{-1.1}$ & \nodata &
  \nodata & 24.7$^{+1.2}_{-1.0}$ & \nodata \\
powerlaw & $\Gamma$ &
  \nodata & 1.01$^{+0.06}_{-0.07}$ & \nodata &
  \nodata & 1.10$^{+0.03}_{-0.03}$ & \nodata &
  \nodata & 1.14$^{+0.03}_{-0.03}$ & \nodata \\
powerlaw & norm &
  \nodata & 0.06$^{+0.01}_{-0.01}$ & \nodata &
  \nodata & 0.53$^{+0.04}_{-0.03}$ & \nodata &
  \nodata & 0.90$^{+0.06}_{-0.05}$ & \nodata \\
CompTT  & $T_0$ (keV) &
  \nodata & \nodata & 0.06$^{+0.01}_{-0.01}$ &
  \nodata & \nodata & 0.09$^{+0.01}_{-0.01}$ &
  \nodata & \nodata & 0.25$^{+0.02}_{-0.02}$ \\
CompTT  & $kT$ (keV) &
  \nodata & \nodata & 12.4$^{+0.1}_{-0.1}$ &
  \nodata & \nodata & 12.1$^{+0.1}_{-0.1}$ &
  \nodata & \nodata & 12.0$^{+0.1}_{-0.2}$ \\
CompTT  & $\tau$ &
  \nodata & \nodata & 3.74$^{+0.06}_{-0.06}$ &
  \nodata & \nodata & 3.38$^{+0.01}_{-0.01}$ &
  \nodata & \nodata & 3.18$^{+0.06}_{-0.07}$ \\
CompTT  & norm ($10^{-2}$) &
  \nodata & \nodata & 5.3$^{+0.2}_{-0.2}$ &
  \nodata & \nodata & 34.5$^{+0.9}_{-1.1}$ &
  \nodata & \nodata & 32.2$^{+2.0}_{-1.3}$ \\
 & $\chi^2_{\rm red}$/dof &
  0.89/309 & 0.89/308 & 0.92/308 &
  0.90/309 & 0.90/308 & 1.14/308 &
  0.91/309 & 0.91/308 & 1.18/308 \\
\enddata
\tablecomments{
Uncertainties are quoted at the 90\% confidence level and are derived from
MCMC chains of length 20\,000. For each observation, three spectral models are applied:
Model~1: {\tt tbabs}~$\times$~({\tt bbodyrad} + {\tt cutoffpl});
Model~2: {\tt tbabs}~$\times$~({\tt bbodyrad} + {\tt highecut}~$\times$~{\tt powerlaw});
Model~3: {\tt tbabs}~$\times$~({\tt bbodyrad} + {\tt CompTT}). The energy range used for spectral fitting is 2--100~keV.
}
\label{table2}
\end{deluxetable*}

In the Insight--HXMT--only results (Figure~\ref{fig2}(a)), we detect a high--temperature blackbody component with \(kT\sim3\,\mathrm{keV}\) and radius \(R\sim1\,\mathrm{km}\), consistent with soft X--ray emission from a hotspot \citep{2020ApJ...902...18K,2022arXiv220414185M}. The blackbody radius was derived from the model normalization as  
\begin{equation}
R_{\rm km} = \sqrt{\mathrm{Norm_{bb}}}\, D_{10},
\end{equation}
where \(\mathrm{Norm_{bb}} = R_{\rm km}^2 / D_{10}^2\). Here \(R_{\rm km}\) is the source radius in km, and \(D_{10}\) is the source distance in units of 10\,kpc. For Swift~J0243.6+6124, we adopt a distance of 5.2\,kpc \citep{2018AJ....156...58B}, corresponding to \(D_{10}=0.52\). The joint fit yields consistent results. In both datasets, the blackbody radius--luminosity relation exhibits a clear break, where the correlation changes from positive to negative.  

To quantitatively characterize this break, we modeled the radius–luminosity relation with a continuous broken--linear function of the form
\begin{equation}
y =
\begin{cases}
y_b + m_1 (x - x_b), & x \leq x_b, \\
y_b + m_2 (x - x_b), & x > x_b,
\end{cases}
\end{equation}
where \(m_1\) and \(m_2\) denote the slopes below and above the break luminosity \(x_b\), and \(y_b\) is the model value at the break point. We excluded data points above \(8.8\times10^{37}\,\mathrm{erg\,s^{-1}}\)( \(2-150\,\mathrm{keV},\,5.2\,\mathrm{kpc}\)), where an additional transition has been reported \citep{2020ApJ...902...18K,2022arXiv220414185M}. 

We use uniform box priors to constrain the parameters of the broken--line model, where $x_b \in [3\times10^{37},\,8\times10^{37}]~\mathrm{erg\,s^{-1}}$, 
$y_b \in [-10^3,\,10^3]$, and $m_1, m_2 \in [-10^3,\,10^3]$. We use emcee with 40 walkers, a burn--in of 2000 steps, and 20\,000 production steps. We report the MAP (Maximum A Posteriori) with 90\% credible intervals for all parameters. We incorporate the distance uncertainty as a systematic component of the final error budget (with a distance of $5.2^{+0.3}_{-0.3}\,\mathrm{kpc})$. Figure~\ref{fig3} shows the final uncertainty bands obtained with and without including the distance uncertainty, demonstrating that the uncertainty introduced by the distance error is non-negligible.

For comparison, we also fit a single--line model $M_0$ described as $y = a + b(x - x_0)$, where $x_0=\langle x\rangle$, and compare it with the continuous broken--line model $M_1$. We center the single--line model at $x_0=\langle x\rangle$. To test whether introducing a break (i.e.\ two additional free parameters) is statistically warranted, we compute for each dataset the differences \(\Delta\ln\mathcal{L}\), \(\Delta\mathrm{AICc}\), and \(\Delta\mathrm{BIC}\), where \(\Delta\ln\mathcal{L}\) quantifies the improvement in log-likelihood and \(\Delta\mathrm{AICc}\) and \(\Delta\mathrm{BIC}\) are the corresponding changes in the corrected Akaike and Bayesian information criteria between the broken and single-power-law models. In addition, we perform a Likelihood--Ratio Test with 10\,000 bootstrap iterations under $M_0$, simulating $y_i^{(\mathrm{sim})}\sim\mathcal{N}[\mu_0(x_i),\sigma_i^2]$ using the $M_0$ MLE (Maximum Likelihood Estimate), refitting $M_0$ and $M_1$ to each realization, and forming the empirical null distribution of $\Delta\ln\mathcal{L}$. 

We conducted tests on four datasets, with detailed results presented in Table~\ref{table1}. For clarity, we present the fitting results of Group~1 and Group~2 in Figure~\ref{fig3}, and the corresponding corner plots are shown in Figure~\ref{fig8} of Appendix~\ref{appendix_A}. The results show that the broken--line model performs better across all datasets, as all $\Delta\mathrm{AICc}$ and $\Delta\mathrm{BIC}$ values exceed~10, and no single LRT (Likelihood--Ratio Test) simulation exceeds the observed $\Delta\ln\mathcal{L}$. Comparing four datasets shows that allowing $N_{\rm H}$ to vary has little impact on the break luminosity, whereas including the 0.7--2\,keV data may slightly shift the transition luminosity toward lower value, indicating that the break is robust (The transition luminosities derived from the four datasets are consistent within the uncertainties at the 90\% confidence level, see Table~\ref{table1}). We also examined the evolution of the constant model parameter and found no systematic variation near the transition(see Figure~\ref{fig10} in Appendix~\ref{appendix_A}). 

Taken together, these results confirm that a genuine break exists in the parameter evolution, corresponding to a new transition luminosity of $L_{\mathrm{b}} = 4.46^{+0.53}_{-0.53}\times10^{37}~\mathrm{erg\,s^{-1}}$ (from Group~1). For consistency with previous studies\citep{2020ApJ...902...18K,2020MNRAS.497.5498W}, we adopt the transition luminosity, $L_{\mathrm{t}}\sim 4.5\times10^{37}~\mathrm{erg\,s^{-1}}$, measured in Group~1 throughout this work.
{%
\setlength{\tabcolsep}{3pt}     
\renewcommand{\arraystretch}{1.05} 

\begin{deluxetable}{llccc}
\tablecaption{Spectral Parameters of the 2018 February 02
\label{tab:spec_20180202_single}}
\tabletypesize{\scriptsize}    
\tablewidth{0pt}               
\tablehead{
\colhead{Component} & \colhead{Parameter} &
\colhead{Joint} & \colhead{NICER} & \colhead{Insight-HXMT}
}
\startdata
constant & $C_1$ &
1.0 (frozen) & -- & 1.0 (frozen) \\
constant & $C_2$ &
$0.89^{+0.010}_{-0.010}$ & -- & $0.97^{+0.014}_{-0.0070}$ \\
constant & $C_3$ &
$0.83^{+0.026}_{-0.025}$ & -- & $0.91^{+0.018}_{-0.023}$ \\
Tbabs & $N_{\rm H}$ ($10^{22}\,\mathrm{cm^{-2}}$) &
$1.1^{+0.012}_{-0.012}$ &
$0.93^{+0.018}_{-0.025}$ &
$0.96^{+0.15}_{-0.18}$ \\
bbodyrad & $kT$ (keV) &
$3.0^{+0.059}_{-0.058}$ &
$0.48^{+0.026}_{-0.022}$ &
$3.0^{+0.026}_{-0.066}$ \\
bbodyrad & norm &
$2.9^{+0.16}_{-0.15}$ &
$590^{+65}_{-110}$ &
$2.5^{+0.16}_{-0.14}$ \\
cutoffpl & $\Gamma$ &
$1.0^{+0.018}_{-0.018}$ &
$0.23^{+0.055}_{-0.071}$ &
$1.0^{+0.032}_{-0.024}$ \\
cutoffpl & $E_{\rm cut}$ (keV) &
$25^{+1.0}_{-0.98}$ &
$8.3^{+0.63}_{-0.74}$ &
$25^{+1.1}_{-0.78}$ \\
cutoffpl & norm &
$0.35^{+0.0059}_{-0.0058}$ &
$0.20^{+0.012}_{-0.012}$ &
$0.36^{+0.022}_{-0.017}$ \\
 & $\chi^2_{\rm red}$/dof &
$0.68/350$ &
$0.61/141$ &
$0.82/332$ \\
\enddata
\tablecomments{
Spectra are fitted with model: {\tt TBabs}$\times$({\tt bbodyrad}+{\tt cutoffpl}). Constants are included only when multiple data groups are jointly fitted.
}
\label{table3}
\end{deluxetable}
}%

We selected three observations just below (20180221), near (20180127), and above (20180117) \(L_{\rm t}\) to illustrate the spectral evolution. The spectral comparison across \(L_{\rm t}\) is shown in Figure~\ref{fig4} and Table~\ref{table2}, and the corner plots of the model parameters are presented in Figure~\ref{fig9} of Appendix~\ref{appendix_A} (Taking the results of Group~1 as an example).
For all three spectral models, the evolution of the blackbody component is consistent, showing an increase followed by a decrease as the luminosity rises, which further supports the presence of a break luminosity.

Together, Figures~\ref{fig2}(a) and ~\ref{fig2}(b) show that across \(L_{\rm t}\) the blackbody temperature steadily increases; the radius–luminosity correlation reverses from positive to negative; \(\Gamma\) evolves from nearly constant to positively correlated with luminosity; and the evolution of \(E_{\rm cut}\) slows. In Figure~\ref{fig5}, the grey dashed line marks the photon index around \(L_{\rm t}\). We observe that on either side of \(L_{\rm t}\), both the blackbody area (\(\mathrm{Norm}_{\rm bb}\)) and the cutoff energy (\(E_{\rm cut}\)) exhibit breaks in their correlation with the photon index, and the evolution of \(E_{\rm cut}\) as a function of \(\Gamma\) becomes noticeably slower. Overall, \(E_{\rm cut}\) is positively correlated with \(\Gamma\), which is consistent with the results reported by \citet{2013A&A...551A...1R} for other sources.

Our results also reveal differences between outbursts. As shown in Figure~\ref{fig2}(a), although the overall spectral evolution during the decay phases is broadly similar, the 2018 giant outburst exhibits systematic deviations. With increasing luminosity, the blackbody temperature evolution of the giant outburst diverges from that of the normal outbursts, and its photon index is systematically higher, indicating a softer spectrum. These differences are most apparent when viewed in the context of the full evolutionary sequence and likely reflect distinct accretion environments during the giant outburst.

Giant (Type~II) outbursts are commonly associated with enhanced mass transfer from a perturbed Be-star circumstellar disk, whereas normal (Type~I) outbursts are typically triggered by increased mass transfer near periastron, naturally leading to different accretion environments\citep{2011Ap&SS.332....1R}. In addition, the soft excess observed during the giant outburst may be related to disk-structure effects: under supercritical accretion, strong radiation pressure near the inner disk can drive optically thick outflows \citep{2019ApJ...873...19T,2020MNRAS.491.1857D}, part of which may be recaptured as the accretion rate declines, producing additional soft X-ray emission. Furthermore, the bifurcation in the blackbody temperature–luminosity relation at high $L_X$ between the 2018 and 2023 outbursts may arise because the 2023 outburst peaked at a luminosity close to $L_1\sim8.8\times10^{37}\ \mathrm{erg\ s^{-1}}$, and therefore did not enter the transition associated with the Zone~II accretion regime \citep{2020ApJ...902...18K}.

\section{Discussion} \label{sec:style}\label{sec4}

We performed a detailed spectral analysis of the decay phases of Swift~J0243.6+6124 observed by Insight--HXMT and NICER during three outbursts. The evolution of the spectral parameters reveals a clear transition at \(L_{\rm t}\sim4.5\times10^{37}\,\mathrm{erg\,s^{-1}}\), constrained by the Insight--HXMT and joint-fit spectra, as the NICER-only fits are limited by their narrow energy coverage. This transition can be observed in multiple alternative models (see Table~\ref{table2}) and is robust under different fitting conditions (see Table~\ref{table1}). To clearly illustrate the position of \(L_{\rm t}\) among the known transitions, we list all transition luminosities identified to date across six orders of magnitude in accretion evolution of Swift~J0243.6+6124 (\(L_{\rm{X}}\sim1.0\times10^{34}\) to \(>1\times10^{39}\,\mathrm{erg\,s^{-1}}\)): \(L_{\rm 1}\sim7\times10^{36}\,\mathrm{erg\,s^{-1}}\) \citep{2023MNRAS.522.6115S}, the newly discovered \(L_{\rm t}\sim4.5\times10^{37}\,\mathrm{erg\,s^{-1}}\), and three higher‐luminosity transitions at \(L_3\sim8.8\times10^{37}\), \(L_4\sim2.8\times10^{38}\)\citep{2020MNRAS.491.1857D,2020ApJ...902...18K}, and \(L_5\sim4.5\times10^{38}\,\mathrm{erg\,s^{-1}}\)\citep{2020MNRAS.497.5498W}(All luminosities have been recalculated assuming a distance of \(5.2\,\mathrm{kpc}.)\).

Among these luminosities, only the transitions at $L_3$ and $L_4$ currently show both spectral and timing signatures. Around $L_1$ and $L_5$, detailed spectral studies are lacking, and transitions are indicated only by timing signatures. The transition at $L_{\rm t}$ is identified from spectral signatures, and its timing study is currently in progress. We emphasize that observing so many transitions in a single source is unprecedented (most sources show only one or two\citep{2013A&A...551A...1R}).

Additionally, the luminosity at which \(L_{\rm t}\) appears is particularly peculiar. Insight--HXMT monitoring has established that the critical luminosity of Swift~J0243.6+6124 occurs at a higher value (\(L_4\)), consistent with the 146\,keV CRSF detection \citep{2020ApJ...902...18K,2022ApJ...933L...3K}. If \(L_{\rm t}\) instead corresponds to a subcritical transition, its luminosity would be one to two orders of magnitude higher than in other accretion-powered X-ray pulsars (\citet{2013A&A...551A...1R}, e.g., EXO~2030+375, KS~1947+300, Swift~J1626.6--5156). This indicates that \(L_{\rm t}\) may represents a previously unrecognized subcritical transition.

The physics in the subcritical regime remains uncertain and several mechanisms have been proposed to explain the emergence of transition luminosities below the critical value\citep{1975A&A....42..311B,1982ApJ...257..733L,2012A&A...544A.123B,2015MNRAS.452.1601P,2022ApJ...939...67B}. A possible association is between \(L_{\rm t}\) and the Coulomb luminosity, \(L_{\rm Coul}\). At very low luminosities (\(L_{\rm{X}}\lesssim10^{34\text{--}35}\,\mathrm{erg\,s^{-1}}\)), the gas‐mediated shock forms close to the neutron‐star surface and inflowing matter decelerates almost entirely at the stellar surface. As \(L_{\rm{X}}\) rises toward \(L_{\rm Coul}\), the emission height increases, the shock front lifts above the surface, and Coulomb interactions complete the final deceleration. The minimum luminosity at which Coulomb drag can halt the inflow is estimated as

\begin{equation}
\label{eq:Lcoul}
\begin{split}
L_{\mathrm{Coul}} = {} & 1.17\times10^{37}\,\mathrm{erg\,s^{-1}}
\left(\frac{\Lambda}{0.1}\right)^{-7/12}
\left(\frac{\tau_*}{20}\right)^{7/12}\\
&\left(\frac{M_*}{1.4\,M_\odot}\right)^{11/8}
\times
\left(\frac{R_*}{10\,\mathrm{km}}\right)^{-13/24}
\left(\frac{B_*}{10^{12}\,\mathrm{G}}\right)^{-1/3}
\end{split}
\end{equation}
where \(\Lambda\) is the accretion‐geometry parameter, \(\tau_*\) the Thomson optical depth parameter, and \(M_*\), \(R_*\), and \(B_*\) are the neutron star mass, radius, and surface magnetic field, respectively\citep{1982ApJ...257..733L,2012A&A...544A.123B}.  
\begin{figure}[!htbp]
    \centering
    \includegraphics[width=0.95\columnwidth]{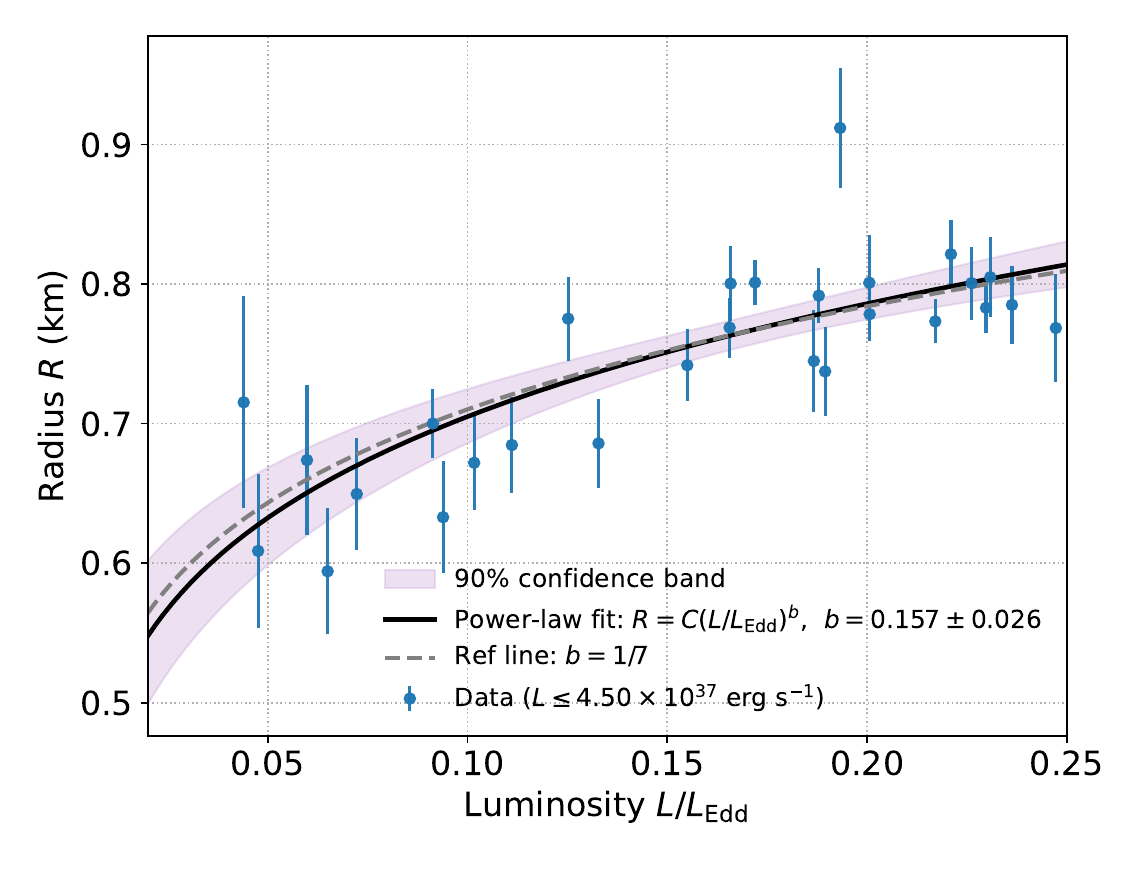}
    \vspace{0.2cm}

    \caption{
    Power-law fit to the radius–luminosity relation using
    \(R = C\,(L/L_{\rm Edd})^{b}\) for the Insight–HXMT-only sample.
    The fit is performed using only data with \(L \le 4.5\times10^{37}\,\mathrm{erg\,s^{-1}}\).
    The purple shaded band marks the 90\% credible interval around the posterior median fit.
    For visual comparison, a gray dashed reference with fixed \(b=1/7\) is overplotted.
    }

    \label{fig6}
\end{figure}

\begin{figure*}[!htbp]
    \centering
    \begin{minipage}{0.4\textwidth}
        \centering
        \includegraphics[width=\textwidth]{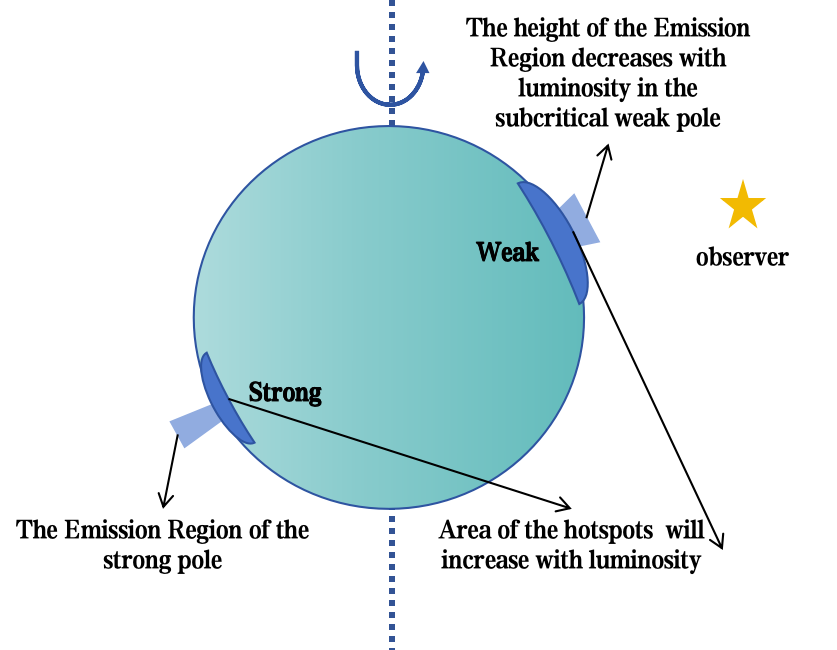}
        \vspace{0.2cm}  
        \makebox[0pt][c]{\textbf{(a) \(L_{\rm X} \lesssim L_{\rm t}\)}}
    \end{minipage}
    \hspace{0.1in}
    \begin{minipage}{0.4\textwidth}
        \centering
        \includegraphics[width=\textwidth]{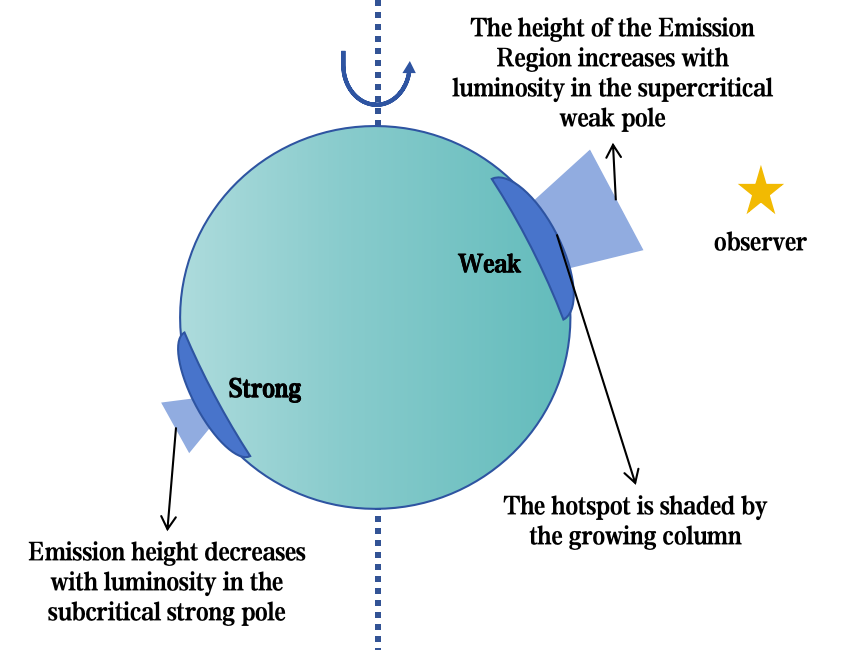}
        \vspace{0.2cm}  
        \makebox[0pt][c]{\textbf{(b) \(L_{\rm X} \gtrsim L_{\rm t}\)}}
    \end{minipage}

    \vspace{0.2cm}  

    \caption{
    Sketch of the two zones defined in this paper during the sub-Eddington regime of Swift~J0243.6+6124. 
    (a) for \(L_{\rm X} \lesssim L_{\rm t}\), the accretion column at the strong magnetic pole is formed, the emission height of the subcritical weak-pole column decreases with luminosity, and the blackbody radius at the polar cap region increases with luminosity; 
    (b) for \(L_{\rm X} \gtrsim L_{\rm t}\), the weak magnetic pole enters the supercritical regime, where the rapidly growing accretion column gradually shades the polar cap region, leading to a decrease in the blackbody radius. The variation of the emission height near the critical luminosity follows the framework discussed by \citet{2012A&A...544A.123B}.
    }

    \label{fig7}
\end{figure*}

If we adopt typical neutron star parameters (\(M_*=1.4\,M_\odot\), \(R_*=10\,\mathrm{km}\)) and set \(\Lambda=0.1\) and \(\tau_*=20\), then the observed \(L_{\rm t}\sim4.5\times10^{37}\,\mathrm{erg\,s^{-1}}\) would imply an implausibly low surface magnetic field of \(B\sim1.7\times10^{10}\,\mathrm{G}\). Even under the most extreme assumptions in the \(L_{\rm Coul}\) formula—namely \(M_*=3\,M_\odot\), \(R_*=15\,\mathrm{km}\), and \(\Lambda=0.66\), matching \(L_{\rm t}\) would require a surface field strength of order \(10^{11}\,\mathrm{G}\), far below previous estimates \citep{2018MNRAS.479L.134T,2020MNRAS.491.1857D,2020ApJ...902...18K,2022ApJ...933L...3K,2023MNRAS.522.6115S}. These calculations strongly suggest that \(L_{\rm t}\) is unlikely to correspond to \(L_{\rm Coul}\).

Theoretical model by \citet{2012A&A...544A.123B} based on a dipole magnetic field predict two transition luminosities, \(L_{\rm Coul}\) and \(L_{\rm crit}\), neither of which matches the observed \(L_{\rm t}\). Moreover, Swift~J0243.6+6124 exhibits as many as five distinct transitions, far exceeding theoretical expectations. We propose that the presence of multipolar magnetic fields may be the underlying cause of \(L_{\rm t}\) and other transitions observed in this source.

A combination of dipole and quadrupole fields naturally produces asymmetric polar‐cap field strengths: one pole is compressed and intensified(strong pole), the other is extended and weakened(weak pole), and if the quadrupole exceeds the dipole locally, the hotspot may become an annular ring\citep{2017ApJ...851..137G,2019MNRAS.490.1774L}. This geometry can give rise to distinct spectral transitions at each pole. In this source, the growth of the column and the emergence of sidewall emission provide an opportunity to view radiation from both poles. Moreover, the existence of multipolar magnetic fields can alter the magnetic field configuration, making the detection of emission from both poles a plausible scenario\citep{2024A&A...691A.123P}.

Under such a magnetic-field configuration, in which the two magnetic poles have different field strengths, the two poles may reach the critical regime at different times or luminosities rather than simultaneously, multiple spectral transitions can naturally arise. We can reconsider the possibility of \(L_{\rm t}\) as the critical luminosity of a certain magnetic pole.

Before \(L_{\rm t}\), a possible spectral hardening is hinted at by comparing the NuSTAR photon index (\(\sim\!1.21\) at relatively low luminosity; Table~2 in \citealt{2019ApJ...873...19T}, OBSID 90401308002) with our below–\(L_{\rm t}\) value (\(\sim\!0.9\)). Thus, the spectrum around \(L_{\rm t}\) most likely follows a hardening–plateau–softening sequence. The hardening and softening can be interpreted by the change of emission height across the critical luminosity, while the plateau represents the transition between the two regimes \citep{2015MNRAS.452.1601P}. We therefore suggest that $L_t$ may represent a critical luminosity,
corresponding to a magnetic field strength of
$B_S \sim 2.8 \times 10^{12}\ \mathrm{G}$,
estimated by substituting typical neutron star parameters into the
critical-luminosity formula:
\begin{equation}
L_{\rm crit} \simeq 1.5 \times 10^{37}\,B_{12}^{16/15}\;\mathrm{erg\,s^{-1}},
\label{eq4}
\end{equation}
where \(B_{12}\) is the surface magnetic field in units of \(10^{12}\,\mathrm{G}\) \citep{2012A&A...544A.123B}.

The previously detected 146~keV cyclotron resonance scattering feature (CRSF) implies the presence of a strong magnetic field near the neutron-star surface, with $B_{\rm N} \sim 1.6 \times 10^{13}\ \mathrm{G}$ \citep{2022ApJ...933L...3K}. When combined with the weaker surface field inferred from the transition luminosity, $B_{\rm S} \sim 2.8 \times 10^{12}\ \mathrm{G}$, this naturally points to an asymmetric magnetic-field configuration between the two magnetic poles. In Appendix~\ref{Appendix_B}, we derive the polar magnetic-field strengths expected from an axisymmetric superposition of dipole and quadrupole components. Substituting the above field estimates into these expressions, we find that reproducing such a two-pole asymmetry requires dipole and quadrupole field strengths of $6.6 \times 10^{12}\ \mathrm{G}$ and $9.4 \times 10^{12}\ \mathrm{G}$, respectively, corresponding to a quadrupole-to-dipole ratio of $q \simeq 1.4$. (After considering the uncertainties of transition luminosity and CRSF line, the ratio is $q=1.42^{+0.06}_{-0.06}$, see Appendix \ref{Appendix_B}.)

Under this magnetic configuration, and further accounting for the measurement
uncertainty of the CRSF energy ($146^{+3}_{-4}$~keV)\citep{2022ApJ...933L...3K}, the magnetic field strength at the stronger pole is $B_{\rm N} = (1.57 \pm 0.04) \times 10^{13}\ \mathrm{G}$, which implies a critical luminosity of $L_{\rm crit} = (2.83 \pm 0.08) \times 10^{38}\ \mathrm{erg\ s^{-1}}$(using equation \ref{eq4}). This value is consistent with the observed $L_4 \sim 2.8 \times 10^{38}\ \mathrm{erg\ s^{-1}}$\citep{2020ApJ...902...18K}, demonstrating that such a magnetic-field configuration can naturally reproduce the observed transition luminosity.

It is worth noting that when the distance from the star increases to $\sim 300$~km (comparable to the magnetospheric radius at high accretion rates for this source\citep{2020MNRAS.491.1857D}), the radial magnetic fields at the two poles have already decayed to nearly equal values, $\simeq 2.2\times10^{8}~\mathrm{G}$. This is equivalent to the field produced at the same radius by a neutron star with a purely dipolar surface polar field of $6.6\times10^{12}~\mathrm{G}$ (see Appendix~\ref{Appendix_B} for the detailed derivation). The magnitude of this effective
dipolar field is consistent with the dipole component inferred from other methods, a few $\times 10^{12}~\mathrm{G}$\citep{2018MNRAS.479L.134T,2020MNRAS.491.1857D,2022MNRAS.516.1601B,2023MNRAS.522.6115S,2023ApJ...952...62I}, while the presence of a
quadrupolar component allows the surface field at one pole to exceed $10^{13}~\mathrm{G}$\citep{2018ApJ...863....9W,2019ApJ...879...61Z,2020ApJ...902...18K,2022ApJ...933L...3K}, thereby reconciling the apparent discrepancies among different magnetic-field estimates.

In addition, when the quadrupole-to-dipole ratio reaches $q \sim 1.4$, a ring-like geometry is expected to form around the weaker magnetic pole \citep{2019MNRAS.490.1774L}, which may cause the evolution of this source to differ from that of typical accreting X-ray pulsars, potentially allow it to reach higher luminosities and weaken the CRSF line. In particular, such a ring geometry can increase the effective length $l_0$ of the accretion-channel cross-section, thereby enhancing the ratio $l_0/d_0$ (where $d_0$ is the thickness of the accretion column) and consequently raising the maximum attainable luminosity (see \citealt{2015MNRAS.454.2539M}, their Eq.~10).

Although a local magnetic field of $B_S \sim 2.8\times10^{12}\,\mathrm{G}$ would place the fundamental cyclotron resonance scattering feature (CRSF) within the Insight--HXMT/NuSTAR band, the detectability of such
a feature depends sensitively on the radiative-transfer conditions rather than on the field strength alone.

Monte Carlo calculations show that the depth and width of CRSFs vary strongly with the line-forming geometry, optical depth, and viewing angle \citep{2007A&A...472..353S}. In particular, the introduction of a quadrupole component leads to larger magnetic-field gradients at the polar region, and the quadrupole-to-dipole ratio of $\sim1.4$ inferred in this work implies an extended, possibly ring-like, weak-pole region \citep{2019MNRAS.490.1774L}. In such a geometry, the observed spectrum originates from a range of emission heights and locations, naturally broadening the CRSF and potentially rendering the CRSF undetectable \citep{2007A&A...472..353S,2014ApJ...781...30N}. In addition, multi-zone photon spawning and the angular dependence of resonant absorption further reduce the line contrast in the continuum \citep{2007A&A...472..353S}.

Turning back to the main transition of the blackbody radius, the photons contributing to this component mainly originate from the spectral region below $\sim15$\,keV.  
Their physical origin remains debated, but it is generally accepted that the softer X-ray photons are primarily produced in the vicinity of the neutron-star surface, either from the polar-cap region or the base of the accretion column\citep{2007ApJ...654..435B,2022ApJ...939...67B}.  
Observationally, the inferred blackbody emitting region has an effective radius of order \(\sim 1 \,km\), and phase-resolved spectroscopy shows a clear modulation of the blackbody parameters with pulse phase, supporting an origin close to the stellar surface\citep{2018MNRAS.474.4432J,2020ApJ...902...18K,2022ApJ...933L...3K}.  
Although a fully self-consistent physical model is still lacking, the blackbody component can be reasonably attributed to emission from the vicinity of the polar caps.  

Under a dipolar magnetic-field assumption, the size of the polar-cap region can be estimated from the Alfvén radius $R_{\rm A}$ as

\begin{align}
R_{\rm cap} &\lesssim R_{\rm NS}\sqrt{\frac{R_{\rm NS}}{R_{\rm A}}}, \label{eq:Rcap}
\end{align}
\begin{align}
R_{\rm A} &= \xi\,\mu^{4/7}\,(2\dot{M})^{-2/7}\,(2GM_{\rm NS})^{-1/7}, \qquad \mu = B_0 R_{\rm NS}^3, \label{eq:RA}
\end{align}

with $\xi \simeq 1$ \citep{1973ApJ...184..271L,2021A&A...651A..12S}.  
Equations~\eqref{eq:Rcap}–\eqref{eq:RA} imply $R_{\rm cap}\propto \dot{M}^{1/7}$.  
We find that the radius–luminosity relation below $L_{\rm t}$ follows this trend well.  
By fitting the pre–$L_{\rm t}$ rise with $R = C\,(L/L_{\rm Edd})^{b}$, we obtain $b\simeq 1/7$, indicating an accretion–rate dependence consistent with that expected for dipolar magnetic fields, even in the possible presence of multipolar components (see Figure.~\ref{fig6}).  
Above the supercritical luminosity, the enhancement of the accretion column may lead to stronger reprocessing of soft X-ray photons from the polar-cap region, thereby causing an apparent decrease of the inferred radius, resembling a self-occultation effect. Assuming that \(L_{\rm t}\) corresponds to the critical luminosity of the weaker magnetic pole, we illustrate in Figure.~\ref{fig7} the schematic geometry of both magnetic poles around \(L_{\rm t}\), highlighting the obscuration of the polar cap region by the accretion column.

We also considered alternative possibilities. Multiple lines of evidence suggest that the transition from a gas-pressure-dominated disk to a radiation-pressure-dominated disk occurs at much higher luminosities, above the Eddington limit \citep{2020MNRAS.491.1857D}. The newly identified transition luminosity is well below the Eddington luminosity and is therefore unlikely to be driven by a structural change of the accretion disk. Moreover, changes in the beaming pattern are generally associated with the onset of the supercritical accretion regime and are expected to be accompanied by significant variations in the pulse profiles \citep{2012A&A...544A.123B,2020ApJ...902...18K}. If $L_{\rm t}$ indeed corresponds to the critical luminosity of the weaker magnetic pole, this scenario can be further examined through pulse-profile evolution.

One plausible scenario is a change in the dominant type of seed photons: at low accretion rates blackbody emission may dominate, whereas at higher rates bremsstrahlung becomes important, and the interplay between these photon sources can produce a characteristic luminosity at which the spectral evolution changes \citep{2022ApJ...939...67B}. This model further predicts a potential spectral hardening process, suggesting that the spectrum becomes harder when bremsstrahlung dominates compared to the regime in which blackbody emission is dominant. However, above \(L_{\rm t}\) we observe spectral softening, with a hardening trend emerging only after \(L_3\). If \(L_{\rm t}\) indeed arises from a change in the seed-photon population, then updated physical models and parameter evolution over a broader luminosity range will be required to verify this interpretation.

\section{Conclusions} \label{sec5}

We have carried out a detailed broadband spectral analysis of Insight–HXMT observations of Swift J0243.6+6124 in its sub‑Eddington regime, complemented by joint fits with contemporaneous NICER data. These combined results reveal a new transition at \(L_{\rm t}\sim4.5\times10^{37}\,\mathrm{erg\,s^{-1}}\) within the sub‑Eddington regime, indicating that this source exhibits unprecedented spectral complexity with at least five intrinsic transition luminosities. This behavior suggests a fascinating multipolar magnetic geometry, in which a weak pole 
(\(B_{\rm weak}\sim2.8\times10^{12}\,\mathrm{G}\)) and a strong pole 
(\(B_{\rm strong}\sim1.6\times10^{13}\,\mathrm{G}\)\citep{2022ApJ...933L...3K}) 
alternately dominate the emission. Furthermore, we found that to give such a magnetic field configuration, both dipole and quadrupole components are comparable. The non-dipole field is equivalent to a pure dipole magnetic field of \(\sim6.6\times10^{12}\,\mathrm{G}\) on the scale of the magnetospheric radius.

\begin{acknowledgments}
This work made primary use of data from the Insight-HXMT mission, a project funded by the China National Space Administration (CNSA) and the Chinese Academy of Sciences (CAS). We thank the High Energy Astrophysics Science Archive Research Center (HEASARC) at NASA's Goddard Space Flight Center for providing the data. This work is supported by the China’s Space Origins Exploration Programthe, National Key R\&D Program of China (2021YFA0718500) and the National Natural Science Foundation of China (NSFC) under Grants No. 12025301 and 12333007. It is also partially supported by the International Partnership Program of the Chinese Academy of Sciences (Grant No. 113111KYSB20190020). Pengju Wang is grateful for the financial support provided by the Sino-German (CSC-DAAD) Postdoc Scholarship Program (57678375)
\end{acknowledgments}

\clearpage
\appendix
\section{SUPPLEMENTARY FIGURES}
\label{appendix_A}
\begin{figure*}[ht!]
    \centering
    \includegraphics[width=0.47\textwidth]{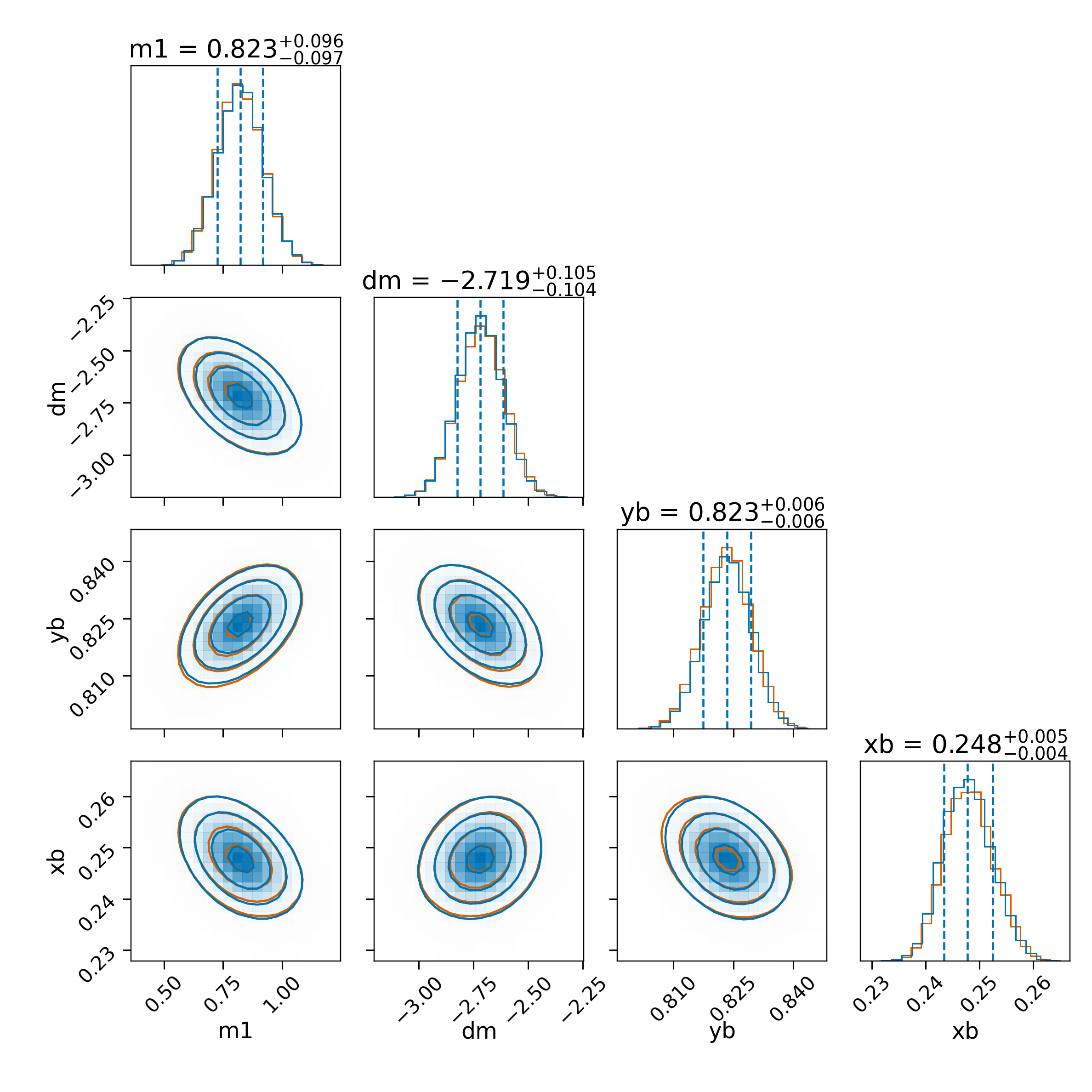}
    \includegraphics[width=0.47\textwidth]{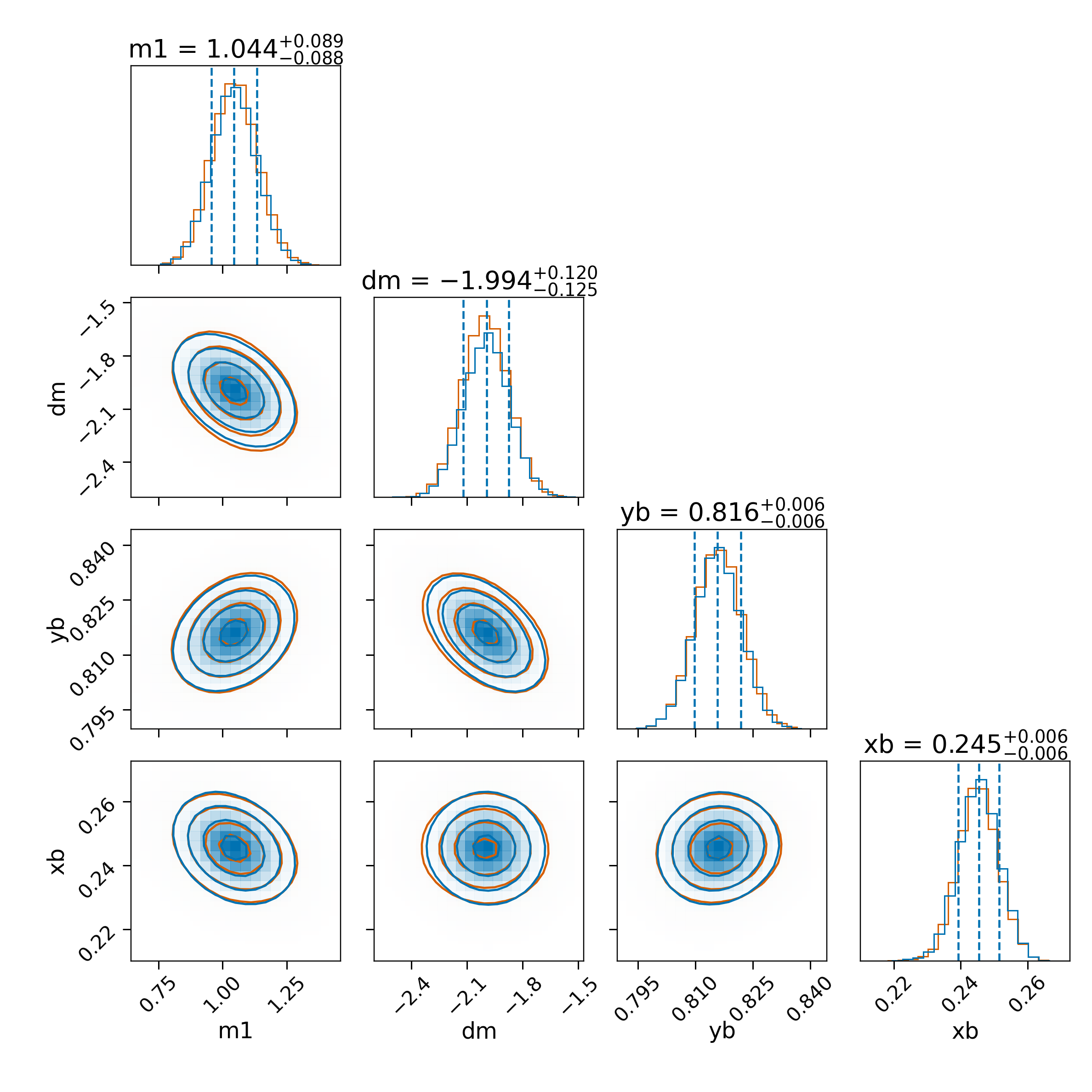}
    \caption{The corner plots for Group~1 and Group~2 in Table~\ref{table1} are shown. Since \(m_{2}\) is parameterized through \(m_{1} + \Delta m\), the corresponding parameter appearing in the plots is \(\Delta m\).
    }
    \label{fig8}
\end{figure*}

\begin{figure*}[ht!]
    \centering
    \includegraphics[width=0.32\textwidth]{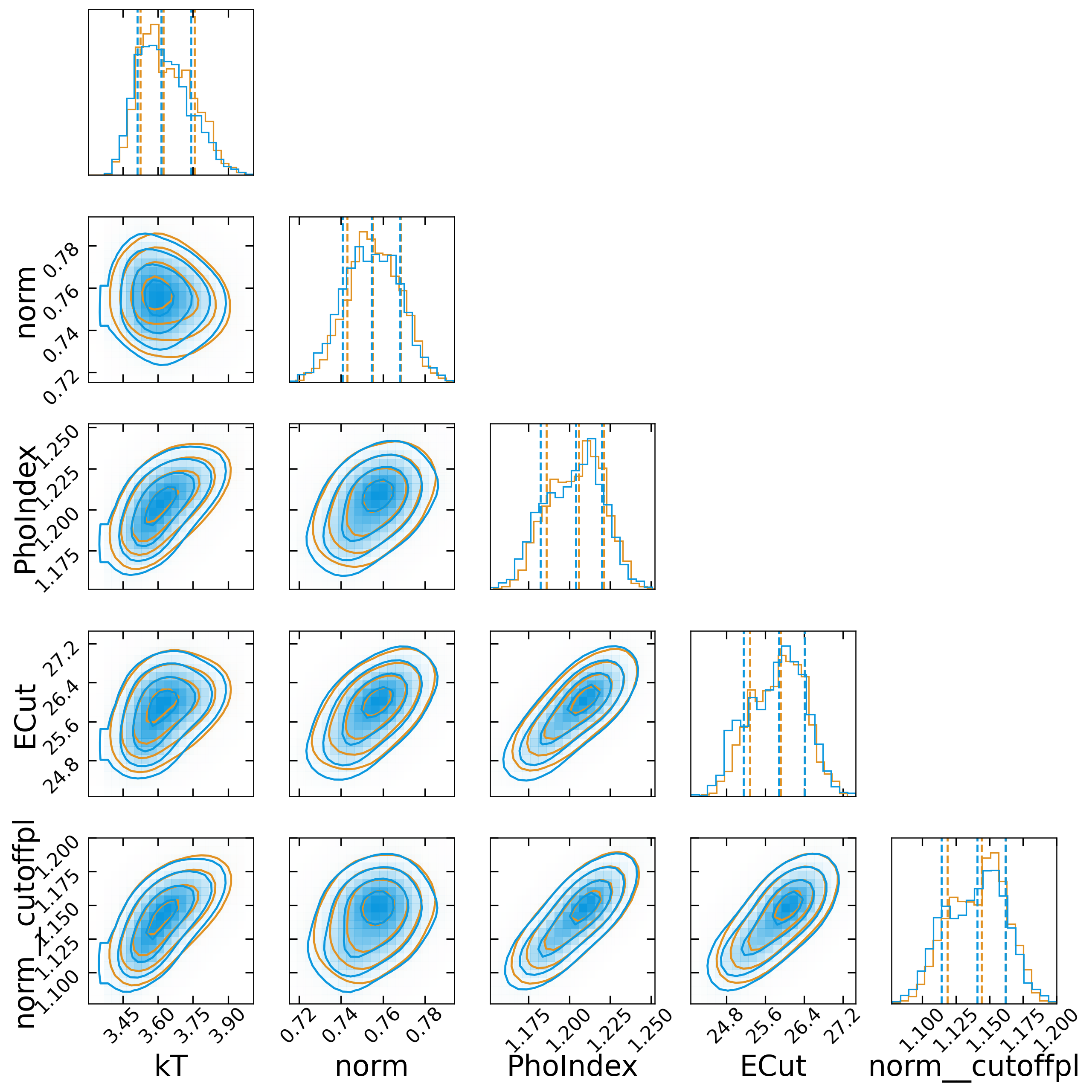}
    \includegraphics[width=0.32\textwidth]{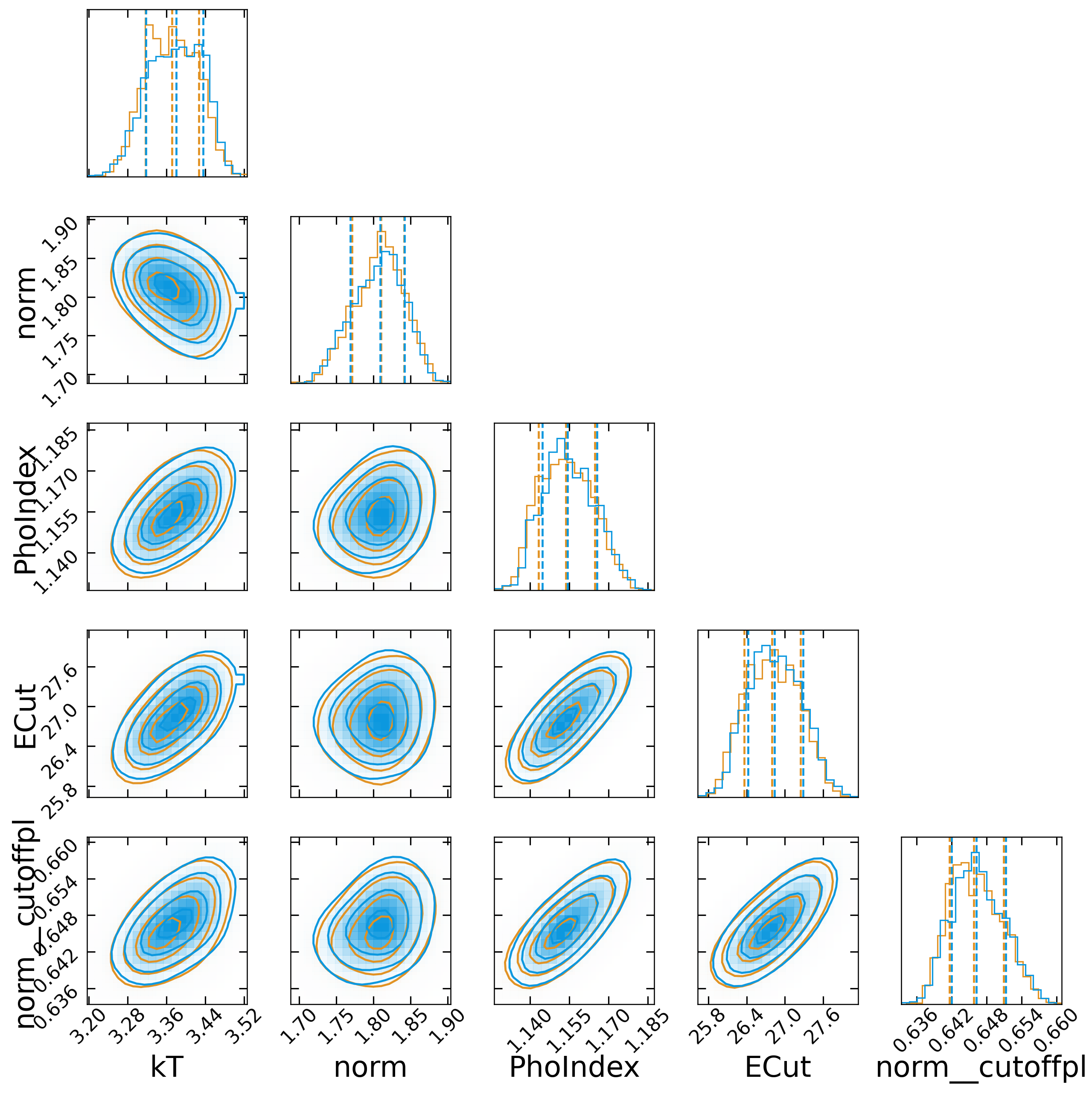}
    \includegraphics[width=0.32\textwidth]{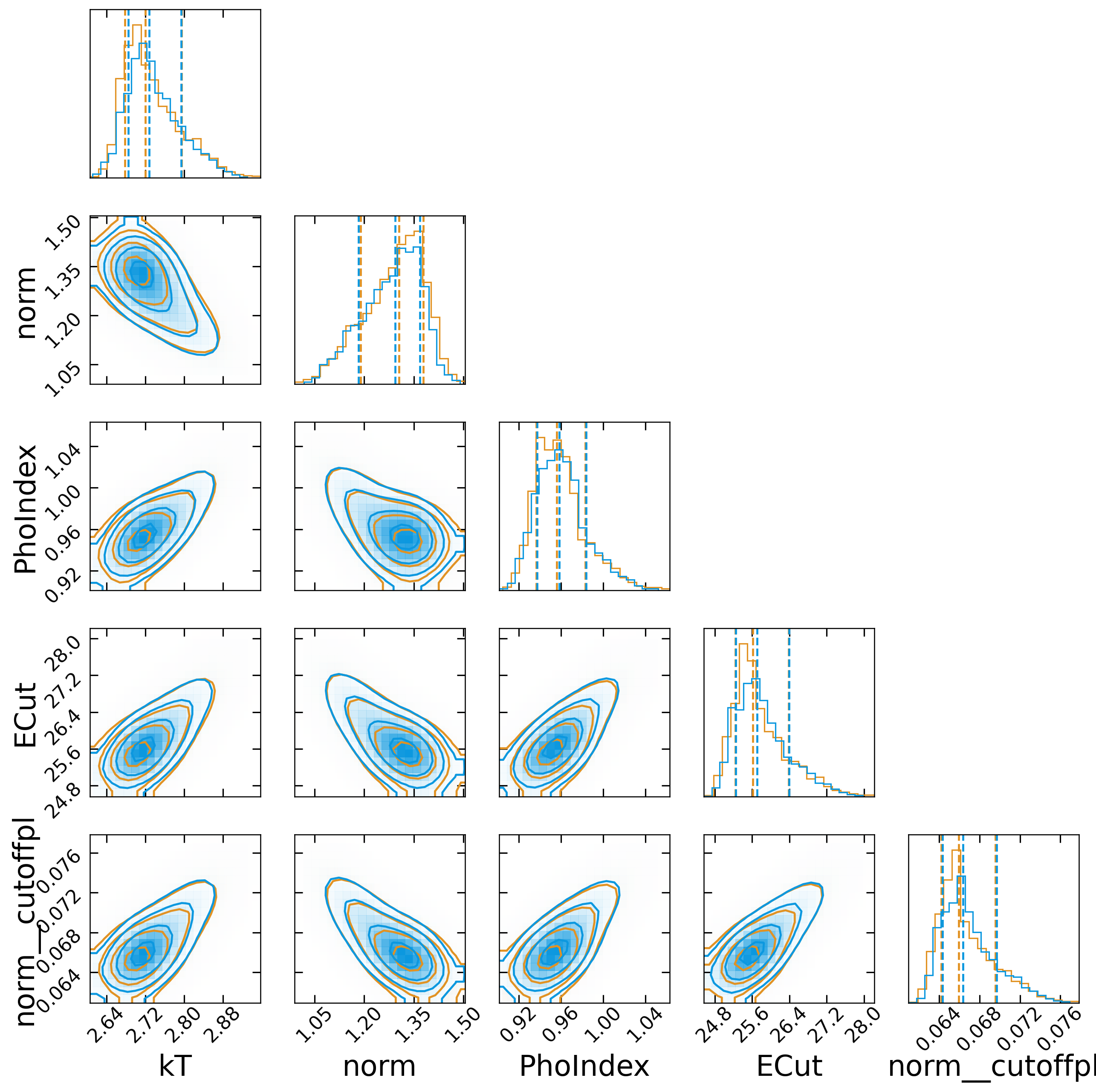}
    \caption{From left to right are the one- and two-dimensional projections of the posterior probability distributions for the three observations with luminosities above $L_{\rm t}$, around $L_{\rm t}$, and below $L_{\rm t}$, respectively. These three observations correspond to those shown in the right panel of Figure~\ref{fig4}: 2018-01-17, 2018-01-27, and 2018-02-21, with luminosities of 
    $8.43\times10^{37}\ {\rm erg\ s^{-1}}$, $6.12\times10^{37}\ {\rm erg\ s^{-1}}$, and $1.17\times10^{37}\ {\rm erg\ s^{-1}}$, respectively. The contours represent the 0.16, 0.50, and 0.84 quantiles obtained from the MCMC analysis of each free spectral parameter. To assess the convergence of the sampling chains, we compare the projections of the posterior distributions derived from the first half (orange) and the second half (blue) of the chains.
    }
    \label{fig9}
\end{figure*}

\begin{figure*}[ht!]
  \centering

  \begin{minipage}[t]{0.42\textwidth}
    \centering
    \includegraphics[width=\linewidth]{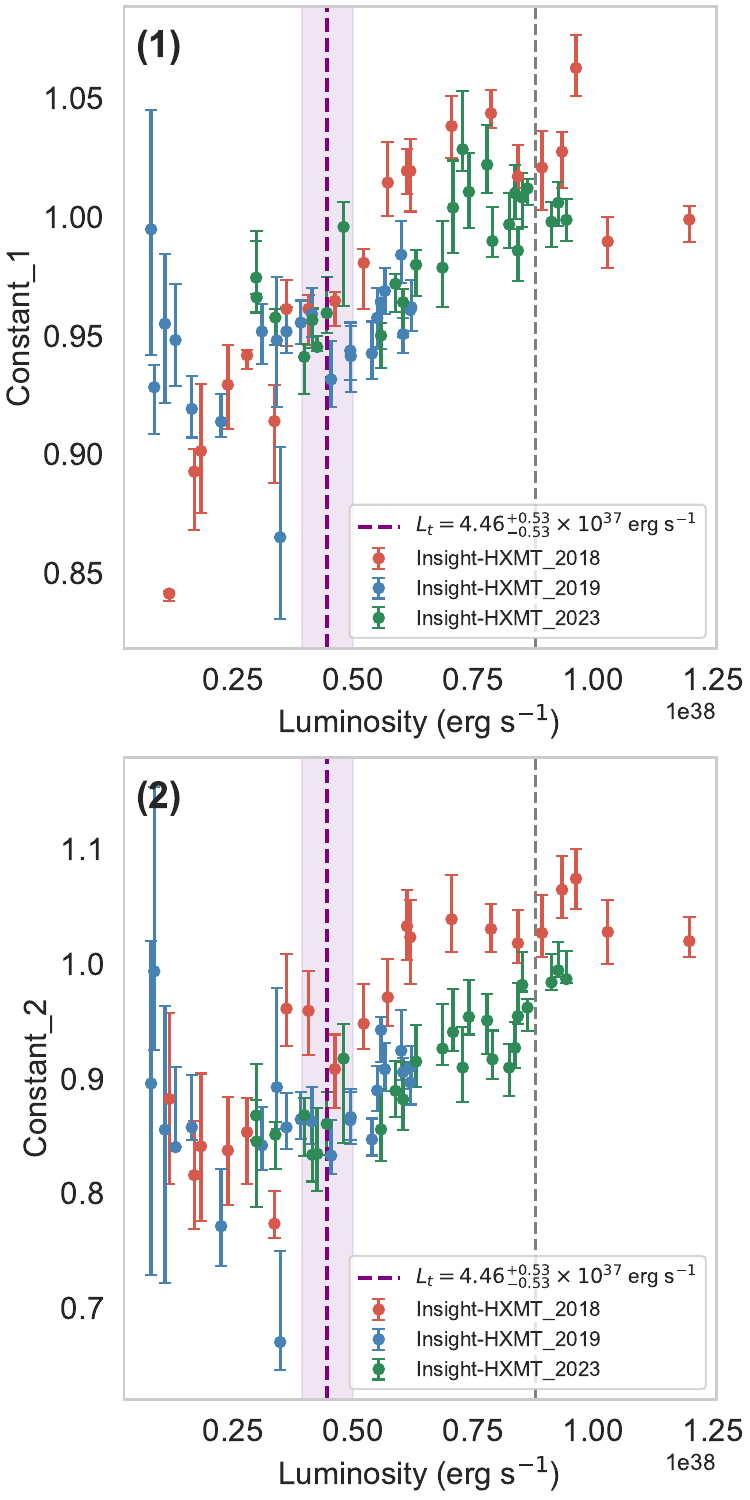}\\
    \textbf{(a) Insight--HXMT}
  \end{minipage}\hfill
  \begin{minipage}[t]{0.42\textwidth}
    \centering
    \includegraphics[width=\linewidth]{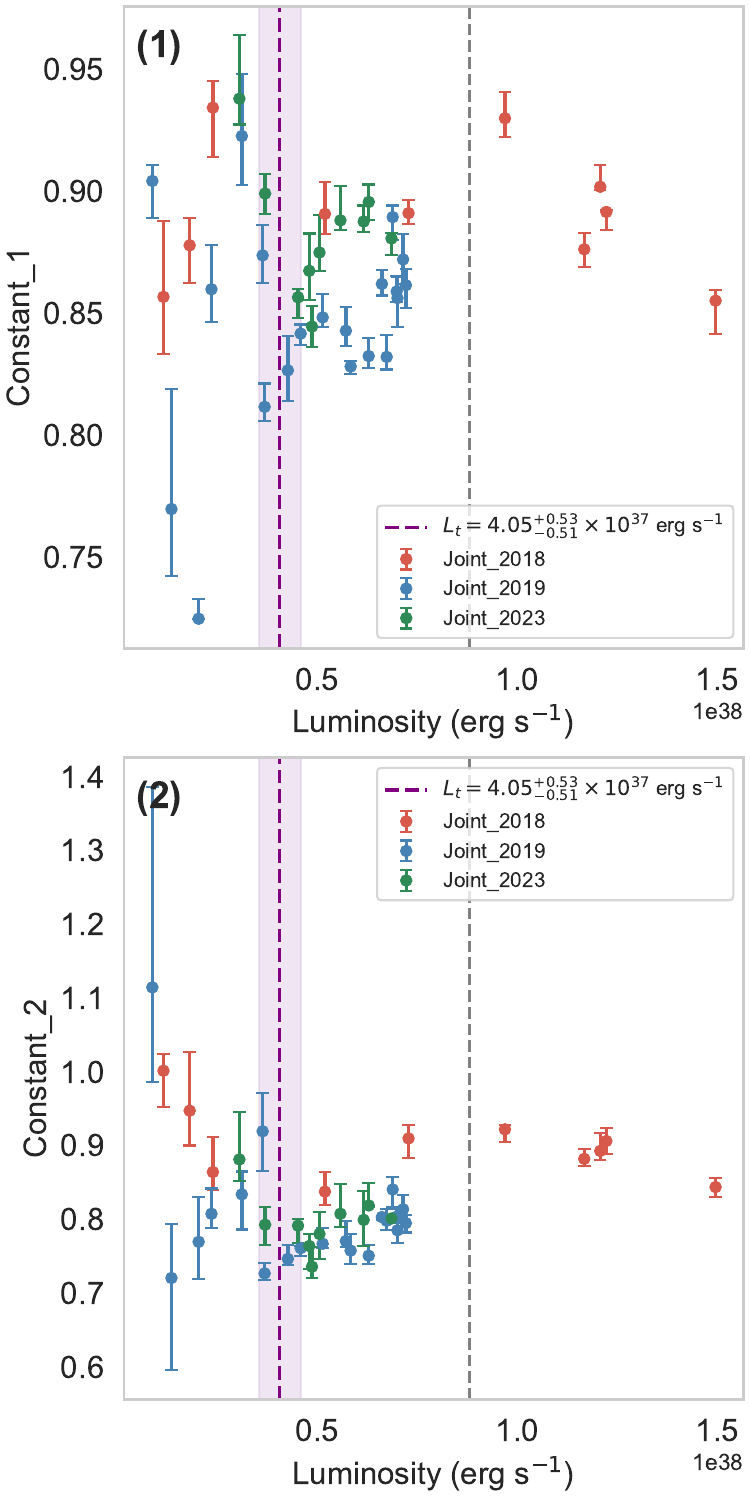}\\
    \textbf{(b) NICER + Insight--HXMT}
  \end{minipage}

  \caption{Supplementary to panels (a) and (b) of Figure~\ref{fig2}, showing the luminosity dependence of the calibration constants for Insight--HXMT-only and joint fits.}
  \label{fig10}
\end{figure*}

\noindent
This Appendix presents supplementary figures supporting the MCMC-based spectral analysis, including corner plots for the grouped fits (Figure~\ref{fig8}), posterior-distribution comparisons for representative observations around $L_{\rm t}$ (Figure~\ref{fig9}), and the luminosity dependence of the cross-calibration constants (Figure~\ref{fig10}).

\FloatBarrier

\section{Multipolar Surface Field from Euler Potentials}
\label{app:euler_multipole}

In an axisymmetric magnetostatic configuration, the magnetic field can be
expressed in terms of Euler potentials as
\begin{equation}
    \mathbf{B} = \nabla\alpha \times \nabla\beta ,
\end{equation}
where we choose
\begin{equation}
    \alpha = \alpha(r,\theta'), \qquad \beta = \phi ,
\end{equation}
and \(\theta'\) is the colatitude measured from the magnetic axis
\citep{2016MNRAS.463.2542G,2017ApJ...851..137G}.

For the ``quadrudipole'' configuration, \citet[][their Eq.~18]{2019MNRAS.490.1774L}
write the Euler potential as
\begin{equation}
    \alpha(r,\theta') =
    B_1 R^2 \frac{R^{>}_1(r)}{R^{>}_1(R)} \sin^2\theta'
  + B_2 R^2 \frac{R^{>}_2(r)}{R^{>}_2(R)} \cos\theta'\,\sin^2\theta' ,
\end{equation}
where \(R\) is the stellar radius, \(B_1\) and \(B_2\) are the surface dipole
and quadrupole coefficients, and \(R^{>}_\ell(r)\) are exterior radial
eigenfunctions.  On the surface (\(r=R\)) one has
\(R^{>}_\ell(r)/R^{>}_\ell(R)=1\), so
\begin{equation}
    \alpha(R,\theta') =
    R^2\left[ B_1 \sin^2\theta' + B_2 \cos\theta'\,\sin^2\theta' \right].
    \label{eq:alpha_surface_simple}
\end{equation}

Using the definition of \(\mathbf{B}\) in spherical coordinates with
\(\beta=\phi\), one finds that for any axisymmetric \(\alpha(r,\theta')\) the
radial component of the magnetic field is
\begin{equation}
    B_r(r,\theta') =
    \frac{1}{r^2\sin\theta'}\,
    \frac{\partial\alpha}{\partial\theta'} .
    \label{eq:Br_from_alpha}
\end{equation}
Differentiating Eq.~\eqref{eq:alpha_surface_simple} with respect to
\(\theta'\) and substituting into Eq.~\eqref{eq:Br_from_alpha} at \(r=R\) gives
\begin{equation}
    B_r(R,\theta') =
    2B_1\cos\theta' + B_2\bigl(3\cos^2\theta' - 1\bigr).
    \label{eq:Br_surface_quadrudipole}
\end{equation}
Evaluating Eq.~\eqref{eq:Br_surface_quadrudipole} at the two magnetic poles
yields
\begin{equation}
    B_r^{\rm N} \equiv B_r(R,0)   = 2(B_1 + B_2),
    \qquad
    B_r^{\rm S} \equiv B_r(R,\pi) = 2(B_2 - B_1),
\end{equation}
where \(B_r^{\rm N}\) and \(B_r^{\rm S}\) are the radial fields at the northern
and southern poles, respectively.  For prescribed polar field strengths
\(B_r^{\rm N}=B_{\rm N}\) and \(B_r^{\rm S}=B_{\rm S}\), these relations
invert to
\begin{equation}
    B_1 = \frac{B_{\rm N} - B_{\rm S}}{4},
    \qquad
    B_2 = \frac{B_{\rm N} + B_{\rm S}}{4}.
    \label{eq:B1B2_from_poles_compact}
\end{equation}
Note that \(B_1\) and \(B_2\) are expansion coefficients in the Euler
potential rather than the surface polar fields of pure dipole or quadrupole
components; for a pure dipole (quadrupole), the corresponding surface polar
field is \(2B_1\) (\(2B_2\)).

As an illustrative example, consider a neutron star with surface polar fields
\begin{equation}
    B_{\rm S} = 2.8\times10^{12}\,{\rm G}, \qquad
    B_{\rm N} = 1.6\times10^{13}\,{\rm G},
\end{equation}
for which Eq.~\eqref{eq:B1B2_from_poles_compact} gives
\begin{equation}
    B_{dipole} = 6.6\times10^{12}\,{\rm G}~(2B_1), \qquad
    B_{quadrupole} = 9.4\times10^{12}\,{\rm G}~(2B_2),
\end{equation}
so that the quadrupole--to--dipole ratio is
\(q \equiv B_2/B_1 \approx 1.42\), i.e.\ the quadrupole and dipole components are of
comparable strength while producing highly asymmetric surface polar fields. Even after taking into account the uncertainties of $L_{\rm t}$ ($4.46^{+0.53}_{-0.53}\times10^{37}\,\mathrm{erg\,s^{-1}}$) and the CRSF energy ($146^{+3}_{-4}\,\mathrm{keV}$)\citep{2022ApJ...933L...3K}, we obtain a dipole field strength of $B_{\rm dip}=6.6^{+0.3}_{-0.2}\times10^{12}\,\mathrm{G}$ and a quadrupole field strength of $B_{\rm quad}=9.4^{+0.2}_{-0.3}\times10^{12}\,\mathrm{G}$, corresponding to a field ratio of $q=1.42^{+0.06}_{-0.06}$. Notably, the ratio remains larger than unity even when the uncertainties are fully considered\citep{2019MNRAS.490.1774L}.

In vacuum, the exterior multipoles scale as
\begin{equation}
    B_1(r) = B_1\left(\frac{R}{r}\right)^3, \qquad
    B_2(r) = B_2\left(\frac{R}{r}\right)^4,
\end{equation}
so that along the magnetic axis
\begin{equation}
    B_r^{\rm N}(r) = 2\,[B_1(r)+B_2(r)], \qquad
    B_r^{\rm S}(r) = 2\,[B_2(r)-B_1(r)].
\end{equation}
For a canonical neutron-star radius \(R_{\rm NS}=10~{\rm km}\), an altitude of
300~km corresponds to \(r\simeq 31\,R_{\rm NS}\), at which we obtain
\begin{equation}
    B_r^{\rm N}(r) \simeq 2.3\times10^{8}~{\rm G}, \qquad
    B_r^{\rm S}(r) \simeq -2.1\times10^{8}~{\rm G},
\end{equation}
so that the magnitudes of the northern and southern polar fields differ by only
\(\sim 10\%\).  Defining an effective polar field at this radius by
\begin{equation}
    B_{\rm eff}(r)
    \equiv \frac{|B_r^{\rm N}(r)| + |B_r^{\rm S}(r)|}{2}
    \simeq 2.2\times10^{8}~{\rm G},
\end{equation}
the surface polar field of an equivalent pure dipole, \(B_{\rm p,dip}\), is
determined from
\begin{equation}
    B_{\rm eff}(r) = B_{\rm p,dip}\left(\frac{R}{r}\right)^3,
\end{equation}
yielding
\begin{equation}
    B_{\rm p,dip}
    \simeq B_{\rm eff}(r)\left(\frac{r}{R}\right)^3
    \simeq 6.6\times10^{12}\,{\rm G},
\end{equation}
which is the surface polar field of a pure dipole that is equivalent, at this
radius, to the multipolar configuration discussed above. Thus, at an altitude
of 300~km, the quadrudipole configuration is effectively indistinguishable from
a pure dipole with \(B_{\rm p,dip}\approx 6.6\times10^{12}\,{\rm G}\).
\label{Appendix_B}


%




\bibliography{sample631}{}
\bibliographystyle{aasjournal}



\end{document}